\documentclass[final,3p,times]{elsarticle}
\usepackage{amssymb,bbold}
\usepackage{amsmath}
\usepackage[utf8x]{inputenc}
\usepackage{natbib}
\usepackage{hyperref}

\hypersetup{
  colorlinks=true,linkcolor=blue,citecolor=blue,
  filecolor=blue,urlcolor=blue,breaklinks=true
}

\biboptions{sort&compress}

\journal{Annals of Physics}

\begin{document}

\begin{frontmatter}
\title{On the spin-1/2 Aharonov-Bohm problem in conical space: bound
  states, scattering and helicity nonconservation}
\author[uepg]{F. M. Andrade}
\ead{fmandrade@uepg.br}
\author[ufma]{E. O. Silva}
\ead{edilbertoo@gmail.com}
\author[uepg]{M. Pereira}
\ead{marciano@uepg.br}

\address[uepg]{
  Departamento de Matem\'{a}tica e Estat\'{i}stica,
  Universidade Estadual de Ponta Grossa,
  84030-900 Ponta Grossa-PR, Brazil
}
\address[ufma]{
  Departamento de F\'{i}sica,
  Universidade Federal do Maranh\~{a}o,
  Campus Universit\'{a}rio do Bacanga,
  65085-580 S\~{a}o Lu\'{i}s-MA, Brazil
}

\begin{abstract}
In this work the bound state and scattering problems for a spin-1/2
particle undergone to an Aharonov-Bohm potential in a conical 
space in the nonrelativistic limit are considered.
The presence of a $\delta$-function singularity, which comes from the
Zeeman spin interaction with the magnetic flux tube, is addressed by the
self-adjoint extension method.
One of the advantages of the present approach is the determination of
the self-adjoint extension parameter in terms of physics of the
problem.
Expressions for the energy bound states, phase-shift and $S$ matrix are
determined in terms of the self-adjoint extension parameter, which is
explicitly determined in terms of the parameters of the problem.
The relation between the bound state and zero modes and the failure of
helicity conservation in the scattering problem and its 
relation with the gyromagnetic ratio $g$ are discussed.
Also, as an application, we consider the spin-1/2 Aharonov-Bohm problem
in conical space plus a two-dimensional isotropic harmonic oscillator.
\end{abstract}


\end{frontmatter}

\section{Introduction}
\label{sec:introduction}

The Aharonov-Bohm (AB) effect \cite{PR.1959.115.485} (first predicted by
Ehrenberg and Siday \cite{PPSB.1949.62.8}) is one of most weird results
of quantum phenomena. The effect reveals that the electromagnetic
potentials, rather than the electric and magnetic fields, are the
fundamental quantities in quantum mechanics. 
The interest in this issue appears in the different contexts, such as
solid-state physics \cite{PRSLA.1931.130.499}, cosmic strings 
\cite{PRD.2013.88.023004,Book.2000.Vilenkin,PRD.2010.82.063515,
PRD.1989.40.1346,BJP.2006.36.141,PRL.2012.108.230404,
AoP.1990.203.392,JMP.1997.38.2553,MPLA.2000.15.253,
MPLA.2004.19.49,AoP.1989.193.142}
$\kappa$-Poincar\'{e}-Hopf algebra
\cite{PLB.1995.359.339,PLB.2013.719.467}, $\delta $-like singularities  
\cite{PRL.1990.64.503,PRD.1993.48.5935,AoP.1996.251.45}, supersymmetry
\cite{AoP.2010.325.2653,PLB.2010.692.51}, condensed matter 
\cite{PRB.2011.84.125306,PRB.2010.82.075316}, Lorentz symmetry violation
\cite{PRD.2011.83.125025}, quantum chromodynamics
\cite{PRD.2010.82.105023}, general relativity \cite{PLB.2011.701.485},
nanophysics \cite{PRL.2010.105.036402}, quantum ring
\cite{PRL.2012.108.106603,PRA.2010.82.022104,PLA.2011.375.2952}, black
hole \cite{PRD.2013.87.125015,PRD.2012.86.125015} and noncommutative
theories \cite{PRD.2011.84.045002,PRD.2002.66.045018}.

In the AB effect of spin-1/2 particles \cite{PRD.1989.40.1346}, besides
the interaction with the magnetic potential, an additional two
dimensional $\delta$-function appears as the mathematical description of
the Zeeman interaction between the spin and the magnetic flux tube
\cite{PRD.1993.48.5935,AoP.1996.251.45}.
This interaction is the basis of the spin-orbit coupling, which
causes a splitting on the energy spectrum of atoms depending on
the spin state.
In Ref. \cite{PRL.1990.64.503} is argued that this
$\delta$-function contribution to the potential can not be
neglected when the system has spin, having shown that changes in
the amplitude and scattering cross section are implied in this case.
The presence of a $\delta$-function potential singularity, turns the
problem more complicated to be solved.
Such kind of point interaction potential can then be addressed by the
self-adjoint extension approach \cite{Book.2004.Albeverio}.
The self-adjoint extension of symmetric operators
\cite{Book.1975.Reed.II} is a very powerful mathematical method and it
can be applied to various systems in relativistic and nonrelativistic
quantum mechanics, supersymmetric quantum mechanics and vortex-like
models.

This paper extends our previous report \cite{PRD.2012.85.041701}
on a general physical regularization method, both in details and
depth.
The method has the advantage of solving problems in relativistic and
nonrelativistic quantum mechanics whose Hamiltonian is singular.
The description of the formalism is based on the works of
Kay-Studer (KS) \cite{CMP.1991.139.103} and Bulla-Gesztesy (BG)
\cite{JMP.1985.26.2520}, both using the self-adjoint
extension method.
The present method is based on the physics of the problem and
one of his particularities is that it gives us the self-adjoint
extension parameter for both bound and scattering scenarios.
Recently, it has been applied for determination of bound states and
scattering matrix for systems with curved surfaces
\cite{JMP.2012.53.122106}, 
quantum deformations \cite{PLB.2013.719.467}, and for AB-like
systems \cite{EPL.2013.101.51005,EPJC.2013.73.2402,JPG.2013.40.075007}.
Here, we address issues which have to do with the existence of
a negative eigenvalue in the spin 1/2 AB spectrum and with the helicity
nonconservation in the scattering. We also add a two-dimensional
isotropic harmonic oscillator in the spin-1/2 AB problem and calculates
the bound states and the self-adjoint extension parameter for this
system.

The paper is organized as follows.
In Sec. \ref{sec:equation_of_motion} we write the Hamiltonian of
the spin-1/2 AB problem and derive the equation of motion that
governs the dynamics of the particle.
In Sec. \ref{sec:selfs} we present the KS and BG self-adjoint
extension methods used in the formulation of the regularization
method proposed here.
The KS method has the advantage of yielding the self-adjoint
extension parameter in terms of the physics of the problem, but
it is not appropriate for dealing with scattering problems; on
the other hand, the BG method is suitable to address both bound
and scattering scenarios, with the disadvantage of allowing
arbitrary self-adjoint extension
parameters.
Further, we also derive the expressions for the energy bound
state, phase shift and the scattering matrix in terms of the
physics of the problem.
By combining the KS and BG methods, a relation between
the self-adjoint extension parameter and the physical parameters
of the problem is found.
In Sec. \ref{sec:applications} we apply the method for the
spin-1/2 AB problem plus a two-dimensional isotropic harmonic
oscillator.
We derive the expression for the particle energy spectrum and
analyze it in the limit case of the vanishing harmonic
oscillator potential recasting the result of usual spin-1/2 AB
problem in conical space.
In Sec. \ref{sec:conclusion} we present a brief conclusion.

\section{The equation of motion}
\label{sec:equation_of_motion}

The idealized situation of a relativistic quantum particle
in the presence of a cosmic string is an example of
gravitational effect of topological origin, where a particle is
transported along a closed curve around the cosmic string
\cite{BJP.2006.36.141}.
This situation corresponds to the gravitational analogue of the
electromagnetic AB effect with the cosmic string replacing the
flux tube
\cite{PRL.2012.108.230404,AoP.1990.203.392,JMP.1997.38.2553,
MPLA.2000.15.253,MPLA.2004.19.49}.
Such effects are of purely topological origin rather than local.
The bound state for the spinless AB effect around a cosmic
string was addressed in \cite{PLA.2007.361.13}.
The authors observed that the self-adjoint extension of the
Hamiltonian of a particle moving around a shielded cosmic string
gives rise to a gravitational analogue of the bound state AB
effect.
Here, our initial proposal is to analyze the spin-1/2 AB
problem in the cosmic string spacetime with an internal magnetic
field.
The cosmic string background is described by the following
metric in cylindrical coordinates ($t,r,\varphi,z$):
\begin{equation}
ds^{2}=-dt^{2}+dr^{2}+\alpha^{2}r^{2}d\varphi^{2}+dz^{2},
\label{eq:metric2}
\end{equation}
with $-\infty<(t,z)<\infty$, $r\geq 0$ and $0\leq\varphi<2\pi$.
The parameter $\alpha$ is related to the linear mass density
$\tilde{m}$ of the string by $\alpha =1-4\tilde{m}$ runs in the
interval $(0,1]$ and corresponds to a deficit angle
$\gamma=2\pi(1-\alpha)$.
The external gravitational field due to a cosmic string may be
approximately described by a commonly called conical
geometry.
Usually, only the case $\alpha <1$ is considered in cosmology,
since $\alpha >1$ corresponds to
a negative mass density cosmic string.
For $\alpha =1,$ the cone
turns into a plane.
The above metric has a cone-like singularity at $r=0$ and the
curvature tensor of this metric, considered as a distribution,
is given by
\begin{equation}
  R_{12}^{12}=R_{1}^{1}=R_{2}^{2}=2\pi
  \left(\frac{ 1-\alpha}{\alpha}\right)
  \delta^2(\mathbf{r}),
  \label{eq:ricci}
\end{equation}
where $\delta^2(\mathbf{r})$ is the two-dimensional
$\delta$-function in flat space \cite{SPD.1977.22.312}.
This implies a two-dimensional conical singularity symmetrical
in the $z$-axis, which characterizes it as a linear defect.

In order to study the dynamics of the particle in a non-flat
spacetime, we should include the spin connection in the
differential operator and define the respective Dirac matrices
in this manifold.
The modified Dirac equation in the curved space reads
\cite{JHEP.2004.2004.16} ($\hbar=c=1$):
\begin{equation}
  \left[
    i\gamma^{\mu}(\partial_{\mu}+\Gamma_{\mu})
    -e\gamma^{\mu}A_{\mu}-M
    \right]
    \Psi=0,
\end{equation}
where $e$ is the charge, $M$ is mass of the particle, $\Psi$ is a
four-component spinorial wave function, and $\Gamma_{\mu}$ is the spin
connection given by
\begin{equation}
  \Gamma_{\mu}=-\frac{1}{4}
  \gamma^{(a)}\gamma^{(b)}e_{(a)}^{\nu}e_{(b)\nu;\mu},
  \label{eq:conn}
\end{equation}
and $\gamma^{\mu}=e_{(a)}^{\mu}(x)\gamma^{(a)}$ are the $\gamma$
matrices in the curved spacetime.
We take the basis tetrad
\cite{JHEP.2004.2004.16,PRD.2008.78.064012,AdP.2010.522.447},
\begin{equation}
e_{(a)}^{\mu}\left( x\right) =\left(
\begin{array}{cccc}
1 & 0 & 0 & 0 \\
0 & \cos \varphi & -\sin \varphi /\alpha r & 0 \\
0 & \sin \varphi & \cos \varphi /\alpha r & 0 \\
0 & 0 & 0 & 1\end{array}\right) ,  \label{eq:tetrad}
\end{equation}
(with $\alpha=1$ giving the flat space-time) satisfying the
condition
\begin{equation}
e_{(a)}^{\mu} e_{(b)}^{\nu}\eta^{(a)(b)}=g^{\mu \nu},
\end{equation}
with $g^{\mu \nu}=\text{diag}(-,+,+,+)$.
For this conical spacetime the spin connection can be expressed
by
\begin{equation}
\gamma^{\mu}\Gamma_{\mu}=-\frac{1-\alpha}{2\alpha r}\gamma^{r},
\label{eq:spconn}
\end{equation}
and
\begin{equation}
  \gamma^{r}=\cos \varphi \gamma^{(1)}+
  \sin \varphi \gamma^{(2)}=
  \left(
    \begin{array}{cc}
      0 & \sigma^{r} \\
      -\sigma^{r} & 0
    \end{array}
  \right) .
\end{equation}
Moreover the $\alpha$ matrices are now written as
\begin{equation}
  \alpha^{i}=e_{(a)}^{i}
  \left(
    \begin{array}{cc}
      0 & \sigma^{(a)} \\
      -\sigma^{(a)} & 0\end{array}\right) =\left(
    \begin{array}{cc}
      0 & \sigma^{i} \\
      \sigma^{i} & 0\end{array}
  \right) ,
\end{equation}
where $\sigma^{i}=(\sigma^{r},\sigma^{\varphi},\sigma^{z})$ are
the Pauli matrices in cylindrical coordinates obtained from the
basis tetrad \eqref{eq:tetrad}.

For the specific tetrad basis used here, the spin connection is
\begin{equation}
\Gamma_{\mu}=( 0,0,\Gamma_{\varphi},0) ,
\label{eq:connec}
\end{equation}
where the nonvanishing element given as
\begin{equation}
\Gamma_{\varphi}=i\frac{(1-\alpha)}{2}\Sigma^{z},
\label{eq:cspconn}
\end{equation}
with $\Sigma^{z}$ being the third component of the spin operator
 $\boldsymbol{\Sigma}=(\Sigma^{r},\Sigma^{\varphi},\Sigma^{z})$,
\begin{equation}
  \Sigma^{r}=
  \left(
    \begin{array}{cc}
      0 & \sigma^{r} \\
      \sigma^{r} & 0
    \end{array}
  \right),
  \;
  \Sigma^{\varphi}=
  \left(
    \begin{array}{cc}
      0 & \sigma^{\varphi}\\
      \sigma^{\varphi} & 0
    \end{array}
  \right),
  \;
  \Sigma^{z}=
  \left(
    \begin{array}{cc}
      \sigma^{z} & 0 \\
      0 & \sigma^{z}
    \end{array}
  \right).
  \label{eq:sigm}
\end{equation}
We are interested in the nonrelativistic limit of the Dirac
equation, so it is convenient to express it in terms of a
Hamiltonian formalism
\begin{equation}
  \hat{H}\psi=\bar{E}\psi,
  \label{eq:2}
\end{equation}
with
\begin{equation}
  \hat{H}=\alpha^{i}\left(-i\nabla_{i}-eA_{i}\right)
  -i\gamma^{0}\gamma^{\mu}\Gamma_{\mu}+\beta M.
  \label{eq:DiracH}
\end{equation}

Exploiting the symmetry under $z$ translations, we can access the
(2+1)-dimensional Dirac equation which follows from the
decoupling of (3+1)-dimensional Dirac equation for the
specialized case where $\partial_{z}=0$ and $A_{z}=0$,
into two uncoupled two-component equations, such as implemented in
Refs.  \cite{PRD.1978.18.2932,NPB.1988.307.909,PRL.1989.62.1071}.
The Dirac equation in $(2+1)$ dimensions reads
\begin{equation}
  \left[
     \beta \mathbf{\gamma} \cdot \boldsymbol{\Pi }+
     \beta M
  \right] \psi =\bar{E}\psi,
\label{eq:gamapi}
\end{equation}
where
\begin{equation}
 \boldsymbol{\Pi} =\frac{1}{i}(\boldsymbol{\nabla}+
 \boldsymbol{\Gamma})-e \mathbf{A},
\end{equation}
is the generalized momentum, $\psi$ is a two-component spinor,
and the $(2+1)$ dimensional $\gamma$ matrices are given in
terms of the Pauli matrices in cylindrical coordinates
\begin{equation}
  \beta=\gamma^{0}=\sigma^{z}, \qquad
  \beta\gamma^{r}=\sigma^{r}, \qquad
  \beta\gamma^{\varphi}=s\sigma^{\varphi},
\end{equation}
where $s$ is twice the spin value, with $s=+1$ for spin ``up''
and $s=-1$ for spin ``down''.

The magnetic flux tube in the background space described by the
metric above considered is related to the magnetic field by
\begin{equation}
  e\mathbf{B}=e
  \boldsymbol{\nabla} \times\mathbf{A}=
 -\frac{\phi}{\alpha}
  \frac{\delta (r)}{r}\mathbf{\hat{z}},
\label{eq:vectorb}
\end{equation}
where $\phi=\Phi/\Phi_{0}$ is the flux parameter with $\Phi_{0}=2\pi/e$,
and the vector potential in the Coulomb gauge is
\begin{equation}
  e\mathbf{A}=
  -\frac{\phi}{\alpha r}
  \hat{\boldsymbol{\varphi}}.
  \label{eq:vectora}
\end{equation}
The choice \eqref{eq:vectorb} also gives the flux tube
coinciding with the cosmic string and the z axis.

The second order equation implied by \eqref{eq:gamapi} is
obtained by applying the matrix operator
$\left[ M+\beta \bar{E}-\mathbf{\gamma }\cdot
 \boldsymbol{\Pi }\right] \beta $.
The result is
\begin{equation}
  (\bar{E}^{2}-M^{2}) \psi =
  \left[
    \boldsymbol{\Pi}_{i}^{2}-es
    \left(
      \boldsymbol{\sigma}\cdot \mathbf{B}
    \right)
  \right]\psi=
  \left[
    \boldsymbol{\Pi}_{i}^{2}+
      \frac{\phi s}{\alpha}\sigma^{z} \frac{\delta(r)}{r}
   \right]\psi,
\end{equation}
In the nonrelativistic limit
\begin{equation}
  \bar{E}=M+E, \qquad M \gg E,
\end{equation}
we have the Schr\"{o}dinger-Pauli equation for $\psi$
\begin{equation}
\left\{
  \left[
    \frac{1}{i}\boldsymbol{\nabla}_{\alpha}+
    \left(
      \frac{1-\alpha}{2\alpha r}\sigma^{z}
      + \frac{\phi}{\alpha r}
    \right)\boldsymbol{\hat{\varphi}}
  \right]^{2}+
  \frac{\phi s}{\alpha}\sigma^{z}\frac{\delta(r)}{r}
  \right\}\psi =k^{2}\psi,
  \label{eq:hpure}
\end{equation}
where $k^{2}=2ME$ and
\begin{equation}
  \nabla_{\alpha}^{2}=
  \frac{\partial^{2}}{\partial r^{2}}+
  \frac{1}{r}\frac{\partial}{\partial r}+
  \frac{1}{\alpha^{2}r^{2}}
  \frac{\partial^{2}}{\partial \varphi^{2}},
\end{equation}
is the Laplacian operator in the conical space.

Before we go on to a calculation of the bound states
and scattering, some remarks on Hamiltonian in \eqref{eq:hpure}
are in order.
If we do not take into account the spin, the resulting
Hamiltonian, in this case, is essentially self-adjoint and
positive definite \cite{PRA.2002.66.032118}.
Therefore, its spectrum is $\mathbb{R}^{+}$, it is
transitionally invariant and there is no bound states.
The introduction of spin changes the situation completely.
The singularity at the origin due to the spin is physically equivalent
to extract this single point from the plane $\mathbb{R}^2$ and 
in this case the translational invariance is lost together with
the self-adjointness.
This fact has impressive consequences in the spectrum of the
system \cite{Book.2008.Oliveira}.
Since we are effectively excluding a portion of space accessible
to the particle we must guarantee that the Hamiltonian is
self-adjoint in the region of the motion, as is necessary for
the generator of time evolution of the wave function.
The most adequate approach for studying this scenario is the
theory of self-adjoint extension of symmetric operators of von
Neumann-Krein
\cite{Book.2004.Albeverio,Book.1975.Reed.II,Book.1993.Akhiezer}.
It yields a family of operators labeled by a real parameter.
We shall see that for all values of this parameter there is an
additional scattering amplitude resulting from the interaction
of the spin with the magnetic flux;
if the parameter is negative there is a bound state with a
negative eigenvalue.
The existence of a negative eigenvalue in the spectrum can be
considered rather unexpected, since the actions of suggest it as a
positive definite operator.
However, the positivity of such an operator does not just depend on
its action, but also depend on its domain.
Indeed, there are several works in the literature which use the
self-adjoint extensions and claim the existence of such a bound
state.
For example, the works of Gerbert \textit{et al.}
\cite{PRD.1989.40.1346,CMP.1989.124.229},
Jackiw \cite{Book.1995.Jackiw} (in this reference an equivalence
between renormalization and self-adjoint extension is
discussed),  Voropaev \textit{et al.} \cite{PLB.1991.267.91},
Bordag \textit{et al.} \cite{PLB.1994.333.238,JPA.1993.26.7637},
Park \textit{et al.} \cite{PRD.1994.50.7715,JMP.1995.36.5453} and
Filgueiras \textit{et al.}
\cite{AoP.2008.323.3150,AoP.2010.325.2529}, to cite few.
In fact, the existence of this negative eigenvalue can be proved
like showen by Albeverio \textit{et al.}
\cite{crll.1987.380.87,Book.2004.Albeverio}.
Now, we can return to our main problem.

Making use of the underlying rotational symmetry expressed by the
fact that $[\hat{H},\hat{J}_{z}]=0$, where
\begin{equation}
\hat{J}_{z}=-i\partial_{\varphi}+\frac{1}{2}\sigma^{z},
\end{equation}
is the total angular momentum operator in the $z$-direction, we
decompose the Hilbert space $\mathfrak{H}=L^{2}(\mathbb{R}^{2})$
with respect to the angular momentum
$\mathfrak{H}=\mathfrak{H}_{r}\otimes\mathfrak{H}_{\varphi}$, where
$\mathfrak{H}_{r}=L^{2}(\mathbb{R}^{+},rdr)$ and
$\mathfrak{H}_{\varphi}=L^{2}(\mathcal{S}^{1},d\varphi)$, with
$\mathcal{S}^{1}$ denoting the unit sphere in $\mathbb{R}^{2}$.
So, it is possible to express the eigenfunctions of the two
dimensional Hamiltonian in terms of the eigenfunctions of
$\hat{J}_{z}$:
\begin{equation}
  \psi(r,\varphi)=
   \left(
    \begin{array}{c}
      f_{m}(r)\; e^{i m \varphi} \\
      g_{m}(r)\; e^{i (m+1)\varphi}
    \end{array}
  \right) ,
  \label{eq:wavef}
\end{equation}
with $m+1/2=\pm 1/2,\pm 3/2, \ldots$, with
$m\in\mathbb{Z}$.
Inserting this into equation \eqref{eq:hpure}, we can extract
the radial equation for $f_{m}(r)$
\begin{equation}
H f_{m}(r)= k^{2} f_{m}(r),
 \label{eq:eigen}
\end{equation}
where
\begin{equation}
H=H_{0}+ \frac{\phi s}{\alpha}\frac{\delta(r)}{r},
\label{eq:hfull}
\end{equation}
and
\begin{equation}
  H_{0}=
    -\frac{d^{2}}{dr^{2}}-\frac{1}{r}\frac{d}{dr}+
    \frac{[m+\phi+(1-\alpha)/2]^{2}}{\alpha^{2}r^{2}}.
  \label{eq:hzero}
\end{equation}
The Hamiltonian in Eq. \eqref{eq:hfull} governs the quantum
dynamics of a spin-1/2 charged particle in the conical
spacetime, with a magnetic field $\mathbf{B}$ along the
z-axis, i.e., a spin-1/2 AB problem in the conical space.
We note that in the case of flat space, $\alpha=1$ (no spin connection),
we recover the radial Hamiltonian for the usual spin-1/2 AB problem in
Refs. \cite{PRL.1990.64.503,PRD.1994.50.7715},
\begin{equation}  \label{eq:habflat}
  -\frac{d^{2}}{dr^{2}}-\frac{1}{r}\frac{d}{dr}+
  \frac{(m+\phi)^2}{r^{2}}+\phi s \frac{\delta(r)}{r}.
\end{equation}
For $\alpha \in (0,1]$ we summarize the possible physical scenarios of
obtaining scattering and bound states in Table \ref{tab:tab1}, based
on the signal of the $\delta$ function coupling constant in
\eqref{eq:hfull}.
Since we have two possibilities for achieving bound states and
scattering, we will focus our attention first on the conditions
giving bound states.
Afterwards, only when we study the scattering problem we
will take into account the other two conditions.

\begin{table}
  \centering
  \begin{tabular}{cccc}
    \hline
    $s$  & $\phi$ & $\phi s/\alpha$ & State \\
    \hline
    $+1$ & $>0$   & $>0$            & Scattering \\
    $-1$ & $<0$   & $>0$            & Scattering \\
    $+1$ & $<0$   & $<0$            & Bound and Scattering \\
    $-1$ & $>0$   & $<0$            & Bound and Scattering\\
    \hline
  \end{tabular}
  \caption{Summary for the physical scenarios based on the signal of
    the $\delta$ coupling constant for $\alpha\in (0,1)$.}
  \label{tab:tab1}
\end{table}

\section{Self-adjoint extensions}
\label{sec:selfs}

In this section we summarize some important concepts and results from
the von Neumann-Krein theory of self-adjoint extensions.
We begin by defining an essentially self-adjoint operator.
An operator $\mathcal{O}$, with domain $\mathcal{D}(\mathcal{O})$, is
said to be essentially self-adjoint if and only if
$\mathcal{D}(\mathcal{O}^{\dagger})=\mathcal{D}(\mathcal{O})$
and $\mathcal{O}^{\dagger}=\mathcal{O}$.
For smooth functions $\xi \in C_{0}^{\infty}(\mathbb{R}^{2})$ with
$\xi(0)=0$, we should have
\begin{equation}
H \xi =H_{0} \xi,
\end{equation}
and hence it is reasonable to interpret
\cite{crll.1987.380.87,JMP.1998.39.47,LMP.1998.43.43} the
Hamiltonian \eqref{eq:hfull} as a self-adjoint extension of
\begin{equation}
H_{0}|_{C_{0}^{\infty}(\mathbb{R}^{2}\setminus \{0\})}.
\end{equation}
Using the unitary operator
$V:L^{2}(\mathbb{R}^{+},rdr)\rightarrow L^{2}(\mathbb{R}^{+},dr)$, given
by $(V\xi)(r)=r^{1/2}\xi(r)$, the operator $H_{0}$ becomes
\begin{equation}
  \tilde{H}_{0}=VH_{0}V^{-1}=
  -\frac{d^{2}}{dr^{2}}+
    \left(\frac{[m+\phi+(1-\alpha)/2]^{2}}{\alpha^{2}}-\frac{1}{4}\right)
    \frac{1}{r^{2}}.
\end{equation}
By standard results, the symmetric radial operator $\tilde{H}_{0}$ is
essentially self-adjoint for $|m+\phi+(1-\alpha)/2|/\alpha\geq 1$.
For those values of $m$ fulfilling $|m+\phi+(1-\alpha)/2|/\alpha<1$ it
is not essentially
self-adjoint, admitting an one-parameter family of self-adjoint
extensions \cite{Book.1975.Reed.II}.
In order to proceed to the self-adjoint extensions of $H_{0}$,
we must find its deficiency subspaces, $N_{\pm}$,
which are defined by
\begin{equation}\label{eq:defspaces}
  N_{\pm} =
  \left\{
    \xi_{\pm} \in \mathcal{D}({H}_{0}^{\dagger}),
    H_{0}^{\dagger}\xi_{\pm} = z_{\pm}\xi_{\pm} ,
    \Im \,z_{\pm} \gtrless 0
  \right\},
\end{equation}
with dimensions $n_{\pm}=\dim {N}_{\pm}$, which are
called deficiency indices of $H_{0}$
\cite{Book.1975.Reed.II}.
A necessary and sufficient condition for $H_{0}$ being
essentially self-adjoint is that $n_{+}=n_{-}=0$.
On the other hand, if $n_{+}=n_{-}\geq 1$ the operator
$H_{0}$ has an infinite number of self-adjoint
extensions parametrized by the unitary operators
$U:N_{+}\rightarrow {N}_{-}$.
Therefore, according to the von Neumann-Krein theory of self-adjoint
extensions, the domain of $H_{0}^{\dagger}$ is given by
\begin{equation}
\mathcal{D}(H_{0}^{\dagger})=
\mathcal{D}(H_{0})\oplus {N}_{+} \oplus {N}_{-}.
\label{eq:dom1}
\end{equation}
One observes that even if the operator is Hermitian
$H_{0}^{\dagger}=H_{0}$, its domains could be different.
The self-adjoint extension approach consists, essentially, in
extending the domain $\mathcal{D}(H_{0})$ to match
$\mathcal{D}(H_{0}^{\dagger})$ in \eqref{eq:dom1}, turning
$H_{0}$ a self-adjoint operator.
We then have
\begin{equation}
\mathcal{D}(H_{\eta,0})=\mathcal{D}(H_{0}^{\dagger})=
\mathcal{D}(H_{0})\oplus {N}_{+} \oplus {N}_{-}.
\end{equation}
where $H_{\eta,0}$ represents the self-adjoint extension of $H_{0}$
parametrized by $\eta \in [0,2\pi)$.

In what follows, to characterize the one parameter family of
self-adjoint extension of $H_{0}$, we will use the KS
\cite{CMP.1991.139.103} and the BG \cite{JMP.1985.26.2520}
approaches, both based on boundary conditions.
In the KS approach, the boundary condition is a match of the
logarithmic derivatives of the zero-energy solutions for
Eq. \eqref{eq:eigen} and the solutions for the problem
$H_{0}$ plus self-adjoint extension.
In the BG approach, the boundary condition is a mathematical
limit allowing divergent solutions for the Hamiltonian
\eqref{eq:hzero} at isolated points, provided they remain square
integrable.

\subsection{KS method}
\label{subsec:KS_method}

In this section, we employ the KS approach to find the bound states for
the Hamiltonian in Eq. \eqref{eq:hfull}.
Following \cite{CMP.1991.139.103}, we temporarily forget the
$\delta$-function potential and find the boundary conditions
allowed for $H_{0}$.
For this intent, we substitute the problem in Eq. \eqref{eq:eigen} by the
eigenvalue equation for $H_{0}$,
\begin{equation}
  H_{0}f_{\rho}=k^{2}f_{\rho},
\label{eq:ideal}
\end{equation}
plus self-adjoint extensions.
Here, $f_{\rho}$ is labeled by the parameter $\rho$ of the self-adjoint
extension, which is related to the behavior of the wave function at the
origin.
In order for the $H_{0}$ to be a self-adjoint operator in
$\mathfrak{H}_{r}$, its domain of definition has to be extended by the
deficiency subspace, which is spanned by the solutions of the eigenvalue
equation (cf. Eq. \eqref{eq:defspaces})
\begin{equation}
H_{0}^{\dagger}f_{\pm}=\pm i k_{0}^{2} f_{\pm},
\label{eq:eigendefs}
\end{equation}
where $k_{0}^{2}\in \mathbb{R}$ is introduced for dimensional
reasons.
Since $H_{0}^{\dagger}=H_{0}$, the only
square integrable functions which are solutions of
Eq. \eqref{eq:eigendefs} are the modified Bessel functions of
second kind,
\begin{equation}
f_{\pm}=K_{[m+\phi+(1-\alpha)/2]/\alpha}( \sqrt{\mp i}k_{0}r),
\end{equation}
with $\Im \sqrt{\pm i}>0$.
These functions are square integrable only in the range
$[m+\phi+(1-\alpha)/2]/\alpha\in(-1,1)$, for which $H_{0}$ is not self-adjoint.
The dimension of such deficiency subspace is
$(n_{+},n_{-})=(1,1)$.
So, we have two situations for $[m+\phi+(1-\alpha)/2]/\alpha$, i.e.,
\begin{align}
\label{eq:jrange}
-1&<[m+\phi+(1-\alpha)/2]/\alpha<0,\\ \nonumber
 0&<[m+\phi+(1-\alpha)/2]/\alpha<1,
\end{align}
and to treat these two situations simultaneously,
it is  more convenient to use
\begin{equation}
  f_{\pm}= K_{|m+\phi+(1-\alpha)/2|/\alpha}(\sqrt{\mp i} k_{0} r).
\end{equation}
Thus, $\mathcal{D}(H_{\rho,0})$ in $L^{2}(\mathbb{R}^{+},rdr)$ is
given by the set of functions \cite{Book.1975.Reed.II}
\begin{equation}
  f_{\rho}(r)=f_{m}(r)+C
  \left[ K_{|m+\phi+(1-\alpha)/2|/\alpha}(\sqrt{-i}k_{0}r)+
    e^{i\rho}K_{|m+\phi+(1-\alpha)/2|/\alpha}(\sqrt{i}k_{0}r)
  \right],
  \label{eq:domain}
\end{equation}
where $f_{m}(r)$, with $f_{m}(0)=\dot{f}_{m}(0)=0$ ($\dot{f}\equiv
df/dr$), is the regular wave function and the parameter
$\rho\in [0,2\pi)$ represents a choice for the boundary condition.
For each $\rho$, we have a possible domain for $H_{0}$ and the
physical situation is the factor that will determine the value of
$\rho$
\cite{AoP.2010.325.2529,AoP.2008.323.3150,PRD.1989.40.1346,
JMP.2012.53.122106}.
Thus, to find a fitting for $\rho$ compatible with the physical
situation, a physically motivated form for the magnetic field  is
preferable for the regularization of the $\delta$-function.
This is accomplished by replacing \eqref{eq:vectora} with
\cite{PRL.1990.64.503,PRL.1990.64.2347,IJMPA.1991.6.3119,
PRD.1993.48.5935}
\begin{equation}
  e\mathbf{A}=
  \left\{
    \begin{array}{lr}
      \displaystyle
      -\frac{\phi}{ \alpha r}
      \mathbf{\hat{\varphi}}, &r>r_{0}\\
      & \\
      0, & r<r_{0}.
  \end{array}
  \right.
\end{equation}
This modification mathematically effects the replacement of
idealized zero thickness filament by one of a finite very small
radius $r_0$ smaller than the Compton wave length
$\lambda_C$ of the electron \cite{PLB.1994.333.238}.
So one makes the replacement
\begin{equation}
\frac{\delta(r)}{r} \to \frac{\delta(r-r_{0})}{r_{0}}.
\label{eq:ushortr0}
\end{equation}
Although the functional structure of
$\delta(r)/r$ and $\delta(r-r_{0})/r_{0}$ are quite different, as
discussed in \cite{PRL.1990.64.503}, we are free to use any form
of potential once that the specific details of the model
\eqref{eq:ushortr0} can be shown to be irrelevant provided that
only the contribution is independent of angle and has no
$\delta$-function contribution at the origin.
It should be remarked that the $\delta(r-r_{0})/r_{0}$ is one
dimensional and well defined contrary to the two dimensional
$\delta(r)/r$.

Now, we are in the position to determine a fitting value for $\rho$.
To do so, we follow \cite{CMP.1991.139.103} and consider the
zero-energy solutions $f_{0}$ and $f_{\rho,0}$ for $H$ with the
regularization in \eqref{eq:ushortr0} and $H_{0}$, respectively, i.e.,
\begin{equation}
  \left[
    -\frac{d^{2}}{dr^{2}}
    -\frac{1}{r}\frac{d}{dr}
    +\frac{[m+\phi+(1-\alpha)/2]^{2}}{\alpha^{2}r^{2}}
    +\frac{\phi s}{\alpha} \frac{\delta(r-r_{0})}{r_{0}}
  \right]
  f_{0}=0,
\label{eq:statictrue}
\end{equation}
\begin{equation}
  \left[
    -\frac{d^{2}}{dr^{2}}
    -\frac{1}{r}\frac{d}{dr}
    +\frac{[m+\phi+(1-\alpha)/2]^{2}}{\alpha^{2}r^{2}}
  \right]
  f_{\rho,0}=0.
\label{eq:rhostatic}
\end{equation}
The value of $\rho$ is determined by the boundary condition
\begin{equation}
\lim_{r_{0}\to 0^{+}}r_{0}\frac{\dot{f}_{0}}{f_{0}}\Big|_{r=r_{0}}=
\lim_{r_{0}\to 0^{+}}r_{0}\frac{\dot{f}_{\rho,0}}{f_{\rho,0}}\Big|_{r=r_{0}}.
\label{eq:logder}
\end{equation}
The left-hand side of this equation can be achieved integrating
\eqref{eq:statictrue} from $0$ to $r_{0}$,
\begin{equation}
  \label{eq:inth0}
  \int_{0}^{r_{0}}\frac{1}{r}\frac{d}{dr}
  \left( r\frac{df_{0}(r)}{dr}\right)rdr=
  \frac{\phi s}{\alpha} \int_{0}^{r_{0}}f_{0}(r)
  \frac{\delta(r-r_{0})}{r_0}rdr +
  \frac{[m+\phi+(1-\alpha)/2]^{2}}{\alpha^{2}}
  \int_{0}^{r_{0}}\frac{f_{0}(r)}{r^{2}}rdr.
\end{equation}
From \eqref{eq:statictrue}, the behavior of $f_{0}$ as $r\to 0$
is $f_{0}\sim r^{|m+\phi+(1-\alpha)/2|/\alpha}$, so we find
\begin{equation}
\int_{0}^{r_{0}}\frac{f_{0}(r)}{r^{2}}rdr\approx
\int_{0}^{r_{0}}r^{|m+\phi+(1-\alpha)/2|/\alpha-1}dr \to 0,
\end{equation}
as $r_{0}\to 0^{+}$.
So, we have
\begin{equation}
\lim_{r_{0}\to 0^{+}}r_{0}\frac{\dot{f}_{0}}{f_{0}}\Big|_{r=r_{0}}
=\frac{\phi s}{\alpha}.
 \label{eq:nrs}
\end{equation}

The right-hand side of Eq. \eqref{eq:logder}
is calculated using the asymptotic representation for
$K_{\nu}(z)$ in the limit $z\rightarrow 0$, given by
\begin{equation}
  K_{\nu}(z)\sim
  \frac{\pi}{2\sin (\pi \nu)}
  \left[
    \frac{z^{-\nu}}{2^{-\nu}\Gamma(1-\nu)}-
    \frac{z^{ \nu}}{2^{ \nu}\Gamma(1+\nu)}
  \right],
  \label{eq:besselasympt}
\end{equation}
in Eq. \eqref{eq:domain}
\footnote{
In Ref. \cite{PRD.2012.85.041701} the expression used for the
asymptotic representation of $K_{\nu}(z)$ it was
$K_{\nu}(z)\sim
\frac{\pi}{2\sin (\pi \nu)}
\left[
\frac{z^{-\nu}}{2^{-\nu}\Gamma(1-\nu)}+
\frac{z^{\nu}}{2^{\nu}\Gamma(1+\nu)}
\right]$,
i.e., the signal of the second term within the brackets must be
minus as in Eq. \eqref{eq:besselasympt}.}.
Thus, we arrive at
\begin{equation}
\lim_{r_{0}\to 0^{+}}r_{0}\frac{\dot{f}_{\rho,0}}{f_{\rho,0}}\Big|_{r=r_{0}}=
\lim_{r_{0}\to 0^{+}}\frac{\dot{\Omega}_{\rho}(r)}{\Omega_{\rho}(r)}\Big|_{r=r_{0}},
  \label{eq:dright}
\end{equation}
where
\begin{align}
 \label{eq:omegas}
  \Omega_{\rho}(r)
  =
  \left[
    \frac
    {\left(\sqrt{-i}k_{0}r\right)^{-|m+\phi+(1-\alpha)/2|/\alpha}}
    {2^{-|m+\phi+(1-\alpha)/2|/\alpha}\Gamma^{(-)}}-
    \frac
    {\left(\sqrt{-i}k_{0}r\right)^{|m+\phi+(1-\alpha)/2|/\alpha}}
    {2^{|m+\phi+(1-\alpha)/2|/\alpha}\Gamma^{(+)}}
  \right]
   +e^{i\rho}
  \left[
    \frac
    {\left(\sqrt{i}k_{0}r\right)^{-|m+\phi+(1-\alpha)/2|/\alpha}}
    {2^{-|m+\phi+(1-\alpha)/2|/\alpha}\Gamma^{(-)}}
    \frac
    {\left(\sqrt{i}k_{0}r\right)^{|m+\phi+(1-\alpha)/2|/\alpha}}
    {2^{|m+\phi+(1-\alpha)/2|/\alpha}\Gamma^{(+)}}
  \right],
\end{align}
where we have introduced the notation
$\Gamma^{(\pm)}=\Gamma\left(1\pm|m+\phi+(1-\alpha)/2|/\alpha\right)$.
Inserting \eqref{eq:nrs} and \eqref{eq:dright} in
\eqref{eq:logder} we obtain
\begin{equation}
  \lim_{r_{0}\to 0^{+}}\frac
  {\dot{\Omega}_{\rho}(r)}
  {\Omega_{\rho}(r)}\Big|_{r=r_{0}}=
  \frac{\phi s}{\alpha},
  \label{eq:saepapprox}
\end{equation}
which gives us the parameter $\rho$ in terms of the physics
of the problem, i.e., the correct behavior of the wave
functions when $r\to 0^{+}$.

We now determine the bound states for $H_{0}$ and using
\eqref{eq:saepapprox} the bound state for $H$ will be determined.
So, we write Eq. \eqref{eq:ideal} for the bound state.
In the present system the energy of a bound state has to be
negative, so that $k$ is a pure imaginary, $k=i\kappa$, with
$\kappa=\sqrt{-2ME_{b}}$, where $E_{b}<0$ is the
bound state energy.
Then, with the substitution $k\to i\kappa$ we have
\begin{equation}
  \left[
    \frac{d^{2}}{dr^{2}}+\frac{1}{r}\frac{d}{dr}-
    \left(\frac{[m+\phi+(1-\alpha)/2]^{2}}{\alpha^{2}r^{2}}+
      \kappa^{2}\right)
  \right] f_{\rho}(r)=0,
  \label{eq:eigenvalue}
\end{equation}
The above equation is the modified Bessel equation whose general
solution is given by
\begin{equation}
  f_{\rho}(r)=K_{|m+\phi+(1-\alpha)/2|/\alpha}
  \left(r\sqrt{-2ME_{b}}\right).
\label{eq:sver}
\end{equation}
Since these solutions belong to $\mathcal{D}(H_{\rho,0})$, it is of
the form \eqref{eq:domain} for some $\rho$ selected from the physics
of the problem.
So, we substitute \eqref{eq:sver} into \eqref{eq:domain} and
compute $\lim_{r_{0}\to 0^{+}}r_{0}{\dot{f}_{\rho}}/{f_{\rho}}|_{r=r_{0}}$
using \eqref{eq:besselasympt}.
After a straightforward calculation, we have the relation
\begin{equation}  \label{eq:derfe}
  \frac
  {|m+\phi+(1-\alpha)/2|/\alpha
    \left[
      r_{0}^{2|m+\phi+(1-\alpha)/2|/\alpha}
      \Gamma^{(-)}
      (-M E_{b})^{|m+\phi+(1-\alpha)/2|/\alpha}+
      2^{|m+\phi+(1-\alpha)/2|/\alpha} \Gamma^{(+)}
    \right]}
  {r_{0}^{2 |m+\phi+(1-\alpha)/2|/\alpha}\Gamma^{(-)}
    (-M E_{b})^{|m+\phi+(1-\alpha)/2|/\alpha}-
    2^{|m+\phi+(1-\alpha)/2|/\alpha} \Gamma^{(+)}}
  =\frac{\phi s}{\alpha}.
\end{equation}
Solving the above equation for $E_{b}$, we find the sought
energy spectrum
\begin{equation}
  E_{b}=-\frac{2}{Mr_{0}^{2}}
  \left[
     \left(
      \frac
      {\phi s+|m+\phi+(1-\alpha)/2|}
      {\phi s-|m+\phi+(1-\alpha)/2|}
    \right)
    \frac
    {\Gamma(1+|m+\phi+(1-\alpha)/2|/\alpha)}
    {\Gamma(1-|m+\phi+(1-\alpha)/2|/\alpha)}
  \right]^{\alpha/|m+\phi+(1-\alpha)/2|}.
\label{eq:energy_KS}
\end{equation}
Notice that there is no arbitrary parameter in the above
equation.
Moreover, to ensure that the energy is a real number, we
must have
\begin{equation}
     \left(
      \frac
      {\phi s+|m+\phi+(1-\alpha)/2|}
      {\phi s-|m+\phi+(1-\alpha)/2|}
    \right)
    \frac
    {\Gamma(1+|m+\phi+(1-\alpha)/2|/\alpha)}
    {\Gamma(1-|m+\phi+(1-\alpha)/2|/\alpha)}
    > 0.
\end{equation}
This inequality is satisfied if $|\phi s|\geq |m+\phi+(1-\alpha)/2|$ and
due to $|m+\phi+(1-\alpha)/2|<1$ it is sufficient to consider
$|\phi s|\geq 1$.
As shown in Table \ref{tab:tab1}, a necessary condition for a $\delta$
function to generate an attractive potential, which is able to support
bound states, is that the coupling constant ($\phi s/\alpha$) must be
negative.
Thus, once that $\alpha\in(0,1]$, the existence of bound states requires
\begin{equation}
  \phi s \leq - 1.
\end{equation}
So, it seems that we must have $\phi s<0$, in such way that the
flux and the spin must be antiparallel, and must have a minimum value
for the $|\phi|$.

\subsection{BG method}

\label{subsec:BG_method}

The KS approach used in the previews section gives us the
energy spectrum in terms of the physics of the problem, but is not
appropriate for dealing with scattering problems.
Furthermore, it selects the value for the parameter $\rho$.
On the other hand, the approach in \cite{JMP.1985.26.2520} is
suitable to address both bound and scattering scenarios, with
the disadvantage of allowing arbitrary self-adjoint extension
parameters.
By comparing the results of these two approaches for bound
states, the  self-adjoint extension parameter can be determined
in terms of the physics of the problem.
Here, all self-adjoint extensions $H_{0,\lambda_{j}}$
of $H_{0}$ are parametrized by the boundary condition
at the origin \cite{JMP.1985.26.2520,Book.2004.Albeverio},
\begin{equation}
\label{eq:bc}
  f^{(0)}=\lambda_{m} f^{(1)},
\end{equation}
with
\begin{eqnarray}
  f^{(0)}&=&\lim_{r\rightarrow 0^{+}}r^{|m+\phi+(1-\alpha)/2|/\alpha} f_{m}(r), \nonumber \\
  f^{(1)}&=&\lim_{r\rightarrow 0^{+}}\frac{1}{r^{|m+\phi+(1-\alpha)/2|/\alpha}}
  \left[
    f_{m}(r)-f^{(0)}\frac{1}{r^{|m+\phi+(1-\alpha)/2|/\alpha}}
  \right],
\end{eqnarray}
where $\lambda_{m}$ is the self-adjoint extension parameter.
In \cite{Book.2004.Albeverio} it is showen that there is a relation
between the self-adjoint extension parameter $\lambda_{m}$  and
the parameter $\rho$ used in the previous section.
The parameter $\rho$ is associated with the mapping of
deficiency subspaces and extend the domain of operator to make
it self-adjoint, being a mathematical parameter.
The self-adjoint extension parameter $\lambda_{m}$ have a
physical interpretation, it represents the scattering length
\cite{Book.2011.Sakurai} of $H_{0,\lambda_{m}}$
\cite{Book.2004.Albeverio}.
For  $\lambda_{m}=0$ we have the free Hamiltonian (without the
$\delta$-function)  with regular wave functions at the origin and
for $\lambda_{m}\neq 0$  the boundary condition in \eqref{eq:bc}
permit a $r^{-|m+\phi+(1-\alpha)/2|/\alpha}$ singularity in the wave
functions at the origin.

\subsubsection{Bound states}
\label{sec:BGbs}

Now we use the BG approach to determine the bound states for $H$ and in 
the end compare with the result obtained with the KS approach.
This allows us to determine the self-adjoint extension parameter in
terms of the physics of the problem.

We begin by rewriting the solutions in another form.
The solutions for
\begin{equation}
  H_{0}f_{m}(r) = k^{2} f_{m}(r),
  \label{eq:hzerofij}
\end{equation}
for $r\neq 0$, taking into account both cases in \eqref{eq:jrange}
simultaneously, can be written in terms of the confluent hypergeometric
function of the first kind  $M(a,b,z)$ as 
\begin{align}
  f_{m}(r) = {} &
  a_{m}e^{-ikr}(2ikr)^{|m+\phi+(1-\alpha)/2|/\alpha}
  M\left(\frac{1}{2}+\frac{|m+\phi+(1-\alpha)/2|}{\alpha},
    1+2\frac{|m+\phi+(1-\alpha)/2|}{\alpha},2ikr \right)\nonumber \\
  & +
  b_{m}e^{-ikr}(2ikr)^{-|m+\phi+(1-\alpha)/2|/\alpha}
  M\left(\frac{1}{2}-\frac{|m+\phi+(1-\alpha)/2|}{\alpha},
    1-2\frac{|m+\phi+(1-\alpha)/2|}{\alpha},2ikr \right),
  \label{eq:general_sol}
\end{align}
where $a_{m}$, $b_{m}$ are the coefficients of the regular and
irregular solutions, respectively.
By implementing Eq. \eqref{eq:general_sol} into the boundary
condition \eqref{eq:bc}, we derive the following relation
between the coefficients $a_{m}$ and $b_{m}$:
\begin{equation}
  \lambda_{m}a_{m}=
  (2ik)^{-2|m+\phi+(1-\alpha)/2|/\alpha}b_{m}
  \bigg(
  1+\frac{\lambda_{m}k^{2}}{4(1-|m+\phi+(1-\alpha)/2|/\alpha)}
  \lim_{r\rightarrow 0^{+}}r^{2-2|m+\phi+(1-\alpha)/2|/\alpha}
\bigg).
 \label{eq:coef_rel_1}
\end{equation}
In the above equation, the coefficient of $b_{m}$ diverges as
$\lim_{r\rightarrow 0^{+}}r^{2-2|m+\phi+(1-\alpha)/2|/\alpha}$, if
$|m+\phi+(1-\alpha)/2|/\alpha\geq 1$.
Thus, $b_{m}$ must be zero for $|m+\phi+(1-\alpha)/2|/\alpha\geq 1$,
and the condition for the occurrence of a singular solution is
$|m+\phi+(1-\alpha)/2|/\alpha<1$.
So, the
presence of an irregular solution stems from the fact the
operator is not self-adjoint for $|m+\phi+(1-\alpha)/2|/\alpha<1$, and
this irregular solution is associated with a self-adjoint extension of
the operator $H_{0}$
\cite{JPA.1995.28.2359,PRA.1992.46.6052}.
In other words, the self-adjoint extension essentially consists in
including irregular solutions in $\mathcal{D}(H_{0})$, which allows to
select an appropriate boundary condition for the problem.

The bound state wave function is obtained with the substitution
$k\rightarrow i\kappa $.
So we have
\begin{align}  \label{eq:bound}
  f_{m}^{b}(r)  = {} &
  a_{m} e^{\kappa r}(-2\kappa r)^{|m+\phi+(1-\alpha)/2|/\alpha}
  M\left(\frac{1}{2}+\frac{|m+\phi+(1-\alpha)/2|}{\alpha},
    1+2\frac{|m+\phi+(1-\alpha)/2|}{\alpha},-2\kappa r\right)\nonumber\\
  &+b_{m}e^{\kappa r}(-2\kappa r)^{-|m+\phi+(1-\alpha)/2|/\alpha}
  M\left(\frac{1}{2}-\frac{|m+\phi+(1-\alpha)/2|}{\alpha},
    1-2\frac{|m+\phi+(1-\alpha)/2|}{\alpha},-2\kappa r\right).
\end{align}
In order to be a bound state  $f_{m}^{b}(r)$
must vanish at large $r$, i.e., it must be normalizable.
By using the asymptotic representation of $M(a,b,z)$ for
$z\rightarrow \infty$,
\begin{equation}
  M(a,b,z)\sim
  \frac{\Gamma(b)}{\Gamma(a)}e^{z}z^{a-b}+
  \frac{\Gamma(b)}{\Gamma(b-a)}(-z)^{-a},
\end{equation}
the normalizability condition yields the relation
\begin{equation}
  \frac{b_{m}}{a_{m}}=-16^{|m+\phi+(1-\alpha)/2|/\alpha}
  \frac{\Gamma^{(+)}}{\Gamma^{(-)}}.
  \label{eq:coef_rel_2}
\end{equation}
From Eq. \eqref{eq:coef_rel_1}, for
$|m+\phi+(1-\alpha)/2|/\alpha<1$ we have
\begin{equation}
  \frac{b_{m}}{a_{m}}=\lambda_{m}(-2\kappa)^{2|m+\phi+(1-\alpha)/2|/\alpha}.
\end{equation}
Combining these two later equations, the bound state energy is determined,
\begin{equation}
  E_{b}=-
  \frac{2}{M}
  \left[
    -\frac{1}{\lambda_{m}}
    \frac
    {\Gamma(1+|m+\phi+(1-\alpha)/2|/\alpha)}
    {\Gamma(1-|m+\phi+(1-\alpha)/2|/\alpha)}
  \right]^{\alpha/|m+\phi+(1-\alpha)/2|}.
  \label{eq:energy_BG}
\end{equation}
This coincides with Eq. (3.13) of Ref. \cite{PRD.1994.50.7715}
for $\alpha=1$, i.e., the spin-1/2 AB problem in flat space.
Also, this coincides with Eq. (26) of
Ref. \cite{JPA.1993.26.7637} for the bound states energy for
particles with an anomalous magnetic moment (with the
replacement $\lambda \to 1/\lambda_{m}$ in that reference).

By comparing Eq. \eqref{eq:energy_BG} with
Eq. \eqref{eq:energy_KS}, we find
\begin{equation}
  \frac{1}{\lambda_{m}}=-\frac{1}{r_{0}^{2|m+\phi+(1-\alpha)/2|/\alpha}}
  \left(
    \frac
    {\phi s+|m+\phi+(1-\alpha)/2|}
    {\phi s-|m+\phi+(1-\alpha)/2|}
  \right).
  \label{eq:lambdaj}
\end{equation}
We have thus attained a relation between the self-adjoint
extension parameter and the physical parameters of the problem.
It should be mentioned that some relations involving the
self-adjoint extension parameter and the $\delta$-function
coupling constant were previously obtained by using Green's
function in Ref. \cite{JMP.1995.36.5453} and the renormalization
technique in Ref. \cite{Book.1995.Jackiw}, being both, however,
deprived from a clear physical interpretation.
Also, in Ref. \cite{JPA.1993.26.7637} a relation between the
self-adjoint extension parameter and the anomaly magnetic moment
was found and it is commented that the dimension of the
self-adjoint extension parameter is $r^{2|m+\phi+(1-\alpha)/2|/\alpha}$
but does not show an explicit relation as found by us in
Eq. \eqref{eq:lambdaj}.

In Ref. \cite{PLB.1994.333.238} the authors comment, based on a
result of Aharonov and Casher \cite{PRA.1979.19.2461} which
states that in a cylindrical magnetic field with flux $\phi$ a
charged particle with magnetic moment and gyromagnetic ratio
$g=2$ possesses $N$ ($N$ being the number of entire flux quanta
in $\phi$) zero modes, i.e., normalizable states with zero
binding energy, any additional attractive force which occurs
for $g>2$ turns the zero modes into bound states.
This $g>2$ value is related with the self-adjoint extension
value, i.e., different values for the self-adjoint extension
parameter corresponds to different values of the $g$
\cite{PLB.1994.333.238}.
The explicit relation between the self-adjoint extension parameter and
the $g$ will be subject of a future work.

Moreover, the bound state wave function is given by
\begin{equation}
  f_{m}^{b}(r)=N_{m}\;
  K_{|m+\phi+(1-\alpha)/2|/\alpha}(-\sqrt{-2ME_{b}}\;r),
  \label{eq:3}
\end{equation}
where $N_{m}$ is a normalization constant and $E_{b}$ is given by
\eqref{eq:energy_BG}.

\subsubsection{Scattering}

Once the bound energy problem has been examined, let us now
analyze the AB scattering scenario.
In this case, the boundary condition is again given by
Eq. \eqref{eq:bc}, but with the replacement
$\lambda_{m}\rightarrow \lambda_{m}^{s}$, where
$\lambda_{m}^{s}$ is the self-adjoint extension parameter for
the scattering problem.
In the scattering analysis it is more convenient to write the
solution for Eq. \eqref{eq:hzerofij} in terms of Bessel
functions
\begin{equation}\label{eq:sol1}
  f_{m}(r)=c_{m}J_{|m+\phi+(1-\alpha)/2|/\alpha}(kr)+
  d_{m}Y_{|m+\phi+(1-\alpha)/2|/\alpha}(kr),
\end{equation}
with $c_{m}$ and $d_{m}$ being constants.
Upon replacing $f_{m}(r)$ in the boundary condition \eqref{eq:bc}, we
obtain
\begin{equation}
  \frac{c_{m}}{d_{m}}=
  \frac
  {\big[\mathcal{B} k^{-|m+\phi+(1-\alpha)/2|/\alpha}-
    \lambda_{m}^{s}(\mathcal{C} k^{|m+\phi+(1-\alpha)/2|/\alpha}+
    \mathcal{B} \mathcal{D}k^{-|m+\phi+(1-\alpha)/2|/\alpha}
    \lim_{r\rightarrow 0^{+}}r^{2-2|m+\phi+(1-\alpha)/2|/\alpha})\big]}
  {\lambda_{m}^{s}\upsilon
    k^{|m+\phi+(1-\alpha)/2|/\alpha}},
\end{equation}
where
\begin{align}
\mathcal{A}  = {} &
\frac{1}{2^{|m+\phi+(1-\alpha)/2|/\alpha}\Gamma^{(+)}},\\
\mathcal{B} ={} &-\frac{2^{|m+\phi+(1-\alpha)/2|/\alpha}
    \Gamma(|m+\phi+(1-\alpha)/2|/\alpha)}{\pi},\\
\mathcal{C} ={} &-\frac{\cos (\pi |m+\phi+(1-\alpha)/2|/\alpha)
    \Gamma(-|m+\phi+(1-\alpha)/2|/\alpha)}{\pi
    2^{|m+\phi+(1-\alpha)/2|/\alpha}},\\
\mathcal{D} ={} & \frac{k^{2}}{4(1-|m+\phi+(1-\alpha)/2|/\alpha)}.
\end{align}

As in the bound state calculation, whenever
$|m+\phi+(1-\alpha)/2|/\alpha<1$, we have
$d_{m}\neq0$; again, this means that there arises the
contribution of the irregular solution $Y_{\nu}(z)$ at the origin when
the operator is not self-adjoint.
Thus, for $|m+\phi+(1-\alpha)/2|/\alpha<1$, we obtain
\begin{equation}
  \frac{c_{m}}{d_{m}}
  =\frac
  {\mathcal{B} k^{-|m+\phi+(1-\alpha)/2|/\alpha}-
    \lambda_{m}^{s}\mathcal{C} k^{|m+\phi+(1-\alpha)/2|/\alpha}}
  {\lambda_{m}^{s}\mathcal{A}
  k^{|m+\phi+(1-\alpha)/2|/\alpha}},
\end{equation}
and by substituting the values of $\mathcal{A}$, $\mathcal{B}$ and
$\mathcal{C}$ into above expression we find
\begin{equation}
d_{m}=-\mu_{m}^{\lambda_{m}^{s}}(k,\phi) c_{m},
\end{equation}
where
\begin{equation}
  \mu_{m}^{\lambda_{m}^{s}}(k,\phi)=
  \frac
  {\lambda_{m}^{s}k^{2|m+\phi+(1-\alpha)/2|/\alpha}\Gamma^{(-)}}
  {\lambda_{m}^{s}k^{2|m+\phi+(1-\alpha)/2|/\alpha}\Gamma^{(-)}
    \cos \left(\pi|m+\phi+(1-\alpha)/2|/\alpha\right)+
    4^{|m+\phi+(1-\alpha)/2|/\alpha}\Gamma^{(+)}}.
  \label{eq:mul}
\end{equation}
Since $\delta$ is a short range potential, it follows that the
behavior of $f_{m}(r)$ for $r\rightarrow \infty $ is given by
\cite{JPA.2010.43.354011}
\begin{equation}
  f_{m}(r)\sim \sqrt{\frac{2}{\pi kr}}
  \cos \left( kr-\frac{|m|\pi}{2}-
  \frac{\pi}{4}+
  \delta_{m}^{{\lambda_{m}^{s}}}(k,\phi)\right) ,
\label{eq:f1asim}
\end{equation}
where $\delta_{m}^{{\lambda_{m}^{s}}}(k,\phi)$  is a
scattering phase shift.
The phase shift is a measure of the argument difference to the
asymptotic behavior of the solution $J_{|m|}(kr)$ of the radial
free equation that is regular at the origin.
By using the asymptotic behavior of $J_{\nu}(z)$ and $Y_{\nu}(z)$ given
in \cite{Book.1972.Abramowitz} in \eqref{eq:sol1}
we obtain
\begin{align}
  \label{eq:scattsol}
  f_{m}(r)
  \sim {} &
  c_{m}\sqrt{\frac{2}{\pi kr}}
  \left[
    \cos\left( kr-\frac{\pi|m+\phi+(1-\alpha)/2|/\alpha}{2}-
      \frac{\pi}{4}\right)\right. \nonumber\\
    & \left. -\mu_{m}^{{\lambda_{m}^{s}}}(k,\phi)
    \sin \left( kr-\frac{\pi |m+\phi+(1-\alpha)/2|/\alpha}{2}-
      \frac{\pi}{4}\right)
  \right] .
\end{align}
By comparing the above expression with Eq. \eqref{eq:f1asim}, we
have
\begin{equation}
  \cos
  \left(
    k r - \frac{\pi |m+\phi+(1-\alpha)/2|/\alpha}{2} -
    \frac{\pi}{4} +\theta_{\lambda_{m}^{s}}
  \right)=
  \cos
  \left( k r- \frac{\pi|m|}{2} - \frac{\pi}{4} +
    \delta_{m}^{{\lambda_{m}^{s}}}(k,{\phi})
  \right),
\end{equation}
with $\theta_{{\lambda_{m}^{s}}}$ given as
\begin{equation}
  \cos \theta_{{\lambda_{m}^{s}}}=c_{m}, \qquad
  \sin \theta_{{\lambda_{m}^{s}}}=c_{m}\; \mu_{m}^{\lambda_{m}^{s}}(k,\phi).
\end{equation}
Therefore, Eq. \eqref{eq:f1asim} is satisfied if
\begin{equation}
  c_{m}=\left[1+
    \left(\mu_{m}^{\lambda_{m}^{s}}(k,\phi)\right)^{2}\right]^{-1/2}.
\end{equation}
Now, comparing the arguments of the cosines above, the following
phase shift is achieved:
\begin{equation}
\delta_{m}^{{\lambda_{m}^{s}}}(k,\phi)=
\Delta_{m}^{AB}(\phi)+\theta_{{\lambda_{m}^{s}}},
\label{eq:phaseshift}
\end{equation}
where
\begin{equation}
\Delta_{m}^{AB}(\phi)=\frac{\pi}{2}(|m|-|m+\phi|),
\end{equation}
is the usual phase shift of the AB scattering and
\begin{equation}
  \theta_{{\lambda_{m}^{s}}}=\arctan {\left(\mu_{m}^{\lambda_{m}^{s}}(k,\phi)\right)}.
\end{equation}
Therefore, the scattering operator
$S_{\phi,m}^{\lambda_{m}^{s}} $ ($S$ matrix) for the
self-adjoint extension is
\begin{equation}
S_{\phi,m}^{\lambda_{m}^{s}}=
e^{2i\delta_{m}^{{\lambda_{m}^{s}}}(k,\phi)}=
e^{2i\Delta_{m}^{AB}(\phi)}e^{2i\theta_{\lambda_{m}^{s}}},
\end{equation}
that is,
\begin{equation}
  S_{\phi,m}^{\lambda_{m}^{s}}=e^{2i\Delta_{m}^{AB}(\phi)}
  \left[
    \frac{1+i\mu_{m}^{\lambda_{m}^{s}}(k,\phi)}{1-i\mu_{m}^{\lambda_{m}^{s}}(k,\phi)}
  \right].
\end{equation}
Using Eq. \eqref{eq:mul}, we have
\begin{equation}
  S_{\phi,m}^{\lambda_{m}^{s}} =
  e^{2i\Delta_{m}^{AB}(\phi)}
  \bigg[
  \frac
  {\lambda_{m}^{s}k^{2|m+\phi+(1-\alpha)/2|/\alpha}\Gamma^{(-)}
    e^{i|m+\phi+(1-\alpha)/2|/\alpha\pi}+4^{|m+\phi+(1-\alpha)/2|/\alpha}\Gamma^{(+)}}
  {\lambda_{m}^{s}k^{2|m+\phi+(1-\alpha)/2|/\alpha}\Gamma^{(-)}
    e^{-i|m+\phi+(1-\alpha)/2|/\alpha\pi}+4^{|m+\phi+(1-\alpha)/2|/\alpha}\Gamma^{(+)}}
  \bigg].
 \label{eq:smatrix}
\end{equation}
Hence, for any value of the self-adjoint extension parameter
$\lambda_{m}^{s}$, there is an additional scattering.
If ${\lambda_{m}^{s}}=0$, we achieve the corresponding result
for the usual AB problem with Dirichlet boundary condition; in
this case, we recover the expression for the scattering matrix
found in Ref. \cite{AoP.1983.146.1},
$S_{\phi ,m}^{0}=e^{2i\Delta_{m}^{AB}(\phi)}$.
If we make ${\lambda_{m}^{s}}=\infty $, we get
$S_{\phi ,m}^{\infty}=e^{2i\Delta_{m}^{AB}(\phi)+2i\pi |m+\phi+(1-\alpha)/2|/\alpha}$.

In accordance with the general theory of scattering, the poles
of the $S$ matrix in the upper half of the complex plane
\cite{PRC.1999.60.34308} determine the positions of the bound
states in the energy scale, Eq. \eqref{eq:energy_BG}.
These poles occur when the denominator of
Eq. \eqref{eq:smatrix} is equal to zero with the replacement
$k\rightarrow i\kappa$.
So, we have
\begin{equation}
  \lambda_{m}^{s}(i\kappa)^{2|m+\phi+(1-\alpha)/2|/\alpha}
  \Gamma^{(-)}e^{-i|m+\phi+(1-\alpha)/2|/\alpha\pi}+
  4^{|m+\phi+(1-\alpha)/2|/\alpha}\Gamma^{(+)}=0.
\end{equation}
Solving this equation for $E_{b}$, we found
\begin{equation}
  E_{b}=-\frac{2}{M}
  \left[-\frac{1}{\lambda_{m}^{s}}
    \frac
    {\Gamma(1+|m+\phi+(1-\alpha)/2|/\alpha)}
    {\Gamma(1-|m+\phi+(1-\alpha)/2|/\alpha)}
  \right]^{1/|m+\phi+(1-\alpha)/2|/\alpha},
  \label{eq:energy_BG-sc}
\end{equation}
for $\lambda_{m}^{s}<0$.
Hence, the poles of the scattering matrix only occurs for
negative values of the self-adjoint extension parameter.
In this latter case, the scattering operator can be expressed in
terms of the bound state energy
\begin{equation}
  S_{\phi,m}^{\lambda_{m}^{s}}=e^{2i\Delta_{m}^{AB}(\phi)}
  \left[
    \frac
    {e^{2 i \pi |m+\phi+(1-\alpha)/2|/\alpha}-
      (\kappa/k)^{2|m+\phi+(1-\alpha)/2|/\alpha}}
    {1-(\kappa/k)^{2|m+\phi+(1-\alpha)/2|/\alpha}}
  \right].
\end{equation}

By comparing Eq. \eqref{eq:energy_BG-sc} above with
Eq. \eqref{eq:energy_BG}, we find $\lambda_{m}^{s}=\lambda_{m}$,
with $\lambda_{m}$ given by Eq. \eqref{eq:lambdaj}, and the
self-adjoint extension parameter for the scattering scenario
being the same one as that for the bound state problem.
This is a very interesting result first discussed in
\cite{PRD.2012.85.041701}.
Thus, we also obtain the phase shift and the scattering matrix
in terms of physics of the problem.

The scattering amplitude $f_{\phi,\alpha}(k,\varphi)$
can be obtained using the standard methods of scattering theory,
namely
\begin{align}
  f_{\phi}^{\alpha}(k,\varphi)
  = {} &
  \frac{1}{\sqrt{2\pi i k}}\sum_{m=-\infty}^{\infty}
  \left(
    S_{m}^{\lambda_{m}}(k,\phi)-1
  \right)
  e^{im\varphi}   \nonumber \\
  ={} &
    \frac{1}{\sqrt{2\pi i k}}
    \left\{
      \sum_{|m+\phi+(1-\alpha)/2|/\alpha\geq 1}
      (e^{2 i \Delta_{m}^{AB}(\phi)}-1)e^{im\varphi}
     +
    \sum_{|m+\phi+(1-\alpha)/2|/\alpha< 1}
    (e^{2 i \Delta_{m}(\alpha)}
    \left[
      \frac
      {1+ i\mu_{m}^{\lambda_{m}}(k,\phi)}
      {1-i\mu_{m}^{\lambda_{m}}(k,\phi)}
    \right]-1)e^{im\varphi}
  \right\}.
  \label{eq:scattamp}
\end{align}
For the special case of $\alpha=1$ (flat space) and $\phi=0$ (zero magnetic
flux) we have $f_{0}^{1}(k,\varphi)=0$, as it should be.
In the above equation we can see that it differs from the usual
AB scattering amplitude off a thin solenoid because
its energy dependence.
As Goldhaber \cite{PRD.1977.16.1815} observed, since the
only length scale in the nonrelativistic problem is set by
$1/k$, it follows that the scattering amplitude
would be a function of the angle alone, multiplied by $1/k$.
This is the manifestation of the helicity conservation.
So, the inevitable failure of helicity conservation expressed in
Eq. \eqref{eq:scattamp} shows that the singularity must lead to
inconsistencies if the Hamiltonian  and the helicity operator,
$\hat{h}=\boldsymbol{\Sigma} \cdot \boldsymbol{\Pi}$,
are treated as well as well-defined operators whose commutation
away from the singularity implies commutation everywhere
\cite{PRD.1977.15.2287,PRL.1983.50.464,PLB.1993.298.63,
NPB.1994.419.323}.
After separation of the variables used in \eqref{eq:wavef}, the
helicity operator is
\begin{equation}
  \hat{h} =
  \left(
    \begin{array}{cc}
      0
      & \displaystyle -i\left(\partial_r+
        \frac{s ([m+\phi+(1-\alpha)/2+1]/\alpha)}{r}\right) \\
      \displaystyle -i\left(\partial_r-
        \frac{s [m+\phi+(1-\alpha)/2]/\alpha}{r}\right) & 0
    \end{array}
  \right).
\end{equation}
This operator suffers from the same issue as the Hamiltonian
operator in the interval $|m+\phi+(1-\alpha)/2|/\alpha<1$, i.e. it is
not self-adjoint \cite{PRD.1994.49.2092,JPA.2001.34.8859}.
Defined on a finite interval $[0,L]$, $\hat{h}$ can be interpreted as a
self-adjoint operator on functions satisfying
$\xi(L)=e^{i\theta}\xi{(0)}$.
However, because the helicity operator must be defined on an infinite
interval $[0,\infty)$, $\hat{h}$ has no self-adjoint extension
\cite{AJP.2001.69.322}, and consequently need not be conserved, and the
helicity can leak at the origin \cite{PRD.1977.16.1815,PRL.1983.50.464}.

As already commented at the end of Section \ref{sec:BGbs}, this result
could be compared with those obtained in Ref. \cite{JPA.1993.26.7637}
where the self-adjoint extension parameter was obtained as a function of
anomaly of the magnetic moment.
In an idealized version of $g-2$ experiment, change in the
helicity in a magnetic field becomes a measure of the departure of the
gyromagnetic ratio of the electron or muon from the Dirac value of
$2e/2M$ \cite{Book.1967.Sakurai}.
For vanishing of $g-2$ there could be no change of helicity even
if the magnetic field were inhomogeneous on a very short length
scale.
So, once again, different values for the self-adjoint extension is
related to different values of the $g$.

\section{The spin 1/2 AB problem plus a two-dimensional
 harmonic oscillator}
\label{sec:applications}

In this section, an application of our method is presented.
We address the spin 1/2 AB problem in conical space plus a two
dimensional harmonic oscillator.
After including the harmonic oscillator (HO) potential and by using the
angular momentum decomposition,
\begin{equation}
  \label{eq:soloc}
  \Phi(r,\varphi)=
  \left(
    \begin{array}{c}
      \chi_{m}(r)\; e^{i m \varphi} \\
      \zeta_{m}(r)\; e^{i(m+1)\varphi}
    \end{array}
\right),
\end{equation}
the radial equation for $\chi_{m}(r)$ becomes
\begin{equation}\label{eq:hoeigen}
H \chi_{m}(r) = k^{2} \chi_{m}(r),
\end{equation}
where
\begin{equation}\label{eq:hho}
  H = H_{0}+ M^2 \omega^{2} r^{2}+
  \frac{\phi s}{\alpha} \frac{\delta(r)}{r}
\end{equation}
with $\omega$ the angular frequency, and $H_{0}$ given by
\eqref{eq:hzero}.
In order to have a more detailed analysis of this problem, we
will first examine the motion of the particle considering two
cases (i) excluding the $r=0$ region and (ii) including the
$r=0$ region afterwards.
At the end, we compare with some results in the literature.

\subsection{Solution of the problem excluding $r=0$ region}

In this case, the Hamiltonian \eqref{eq:hho} does not include
the delta function potential.
By directly solving \eqref{eq:hoeigen} we obtain
\cite{Book.1972.Abramowitz}
\begin{align}\label{eq:ku1}
  \chi_{m}(r)
  = {} & a_{m} (M\omega)^{1/2+[m+\phi+(1-\alpha)/2]/2\alpha}
  r^{[m+\phi+(1-\alpha)/2]/\alpha}
  e^{-M\omega r^{2}/2}
  M\left(d,1+\frac{m+\phi+(1-\alpha)/2}{\alpha},M\omega r^{2}\right)
  \nonumber \\
  & + b_{m}(M\omega)^{{1/2+[m+\phi+(1-\alpha)/2]/2\alpha}}
  r^{[m+\phi+(1-\alpha)/2]/\alpha}
  e^{-{M \omega r^{2}}/{2}}
  U\left(d,1+\frac{m+\phi+(1-\alpha)/2}{\alpha},M\omega r^{2}\right),
\end{align}
where
\begin{equation}
d=\frac{1}{2}\left(1+\frac{m+\phi+(1-\alpha)/2}{\alpha}\right)-
\frac{E}{2\omega},
\end{equation}
$U(a,b,z)$ is the confluent hypergeometric function of the
second kind, and $a_{m}$, $b_{m}$ are constants.
However, as only $M(a,b,z)$ is regular at the origin, it should be imposed
$b_{m}=0$.
Moreover, if $d$ is 0 or a negative integer the series terminates and the
hypergeometric function becomes a polynomial of degree $n$
\cite{Book.1972.Abramowitz}.
This condition guarantees that the confluent hypergeometric function is
regular at the origin, which is essential for the treatment of the
physical system since the region of interest is that around the
flux tube.
Therefore, the series in \eqref{eq:ku1} must converge if we
consider that $d=-n$, $n\in\mathbb{Z}^{*}$, with
$\mathbb{Z}^{*}$ denoting the set of the nonnegative integers.
This condition also guarantees the normalizability of the wave
function.
So, using this condition, we obtain the discrete values for the
energy whose expression is given by
\begin{equation}
  E_{b}=
  \left(
    2n+1+\frac{|m+\phi+(1-\alpha)/2|}{\alpha}
  \right)\omega,
  \qquad n=0,1,2,\ldots.
  \label{eq:en1}
\end{equation}
The  bound state wave function is given by
\begin{equation}
  \label{eq:wf1}
  \chi_{m}^{b}(r) =
  N_{m} \; (M\omega)^{{1/2+[m+\phi+(1-\alpha)/2]/2\alpha}}
  r^{|m+\phi+(1-\alpha)/2|/\alpha}
  e^{-M\omega r^{2}/2}
  M\left(-n,1+\frac{m+\phi+(1-\alpha)/2}{\alpha},M\omega r^{2}\right),
\end{equation}
with $N_{m}$ a normalization constant.
Notice that in Eq. \eqref{eq:en1}, $[m+\phi+(1-\alpha)/2]/\alpha$ can
assume any value.
However, we will see that this condition is no longer true when
we include the $\delta$ function.
Next to study the motion of the particle in all space, including
the $r=0$ region, the self-adjoint extension approach is invoked.

\subsection{Solution including the $r=0$ region}

In this case, the dynamics includes the $\delta$ function.
So, we follow the procedure outlined in Sec. \ref{subsec:KS_method} to
find the bound states for the system.
Like before we need to find all the self-adjoint extension for the
operator $H_{0}+M^{2}\omega^{2}r^{2}$.
So, we substitute the problem in Eq. \eqref{eq:ku1} by
\begin{equation}
  [H_{0}+M^{2}\omega^{2}r^{2}]\chi_{\vartheta}(r)=
  k^{2}\chi_{\vartheta}(r),
\label{eq:autovalor_th}
\end{equation}
plus self-adjoint extensions, with $\chi_{\vartheta}$ labeled by a
parameter $\vartheta$.
The solution to this equation is given in \eqref{eq:ku1}.
However, the only square integrable function is
$U(d,1+[m+\phi+(1-\alpha)/2]/\alpha,M\omega r^{2})$.
Then, this implies that $a_{m}=0$ in Eq. \eqref{eq:ku1}, and
we have
\begin{equation}
  \chi_{\vartheta}(r)=
  (M \omega)^{{1/2+[m+\phi+(1-\alpha)/2]/2\alpha}}
  r^{[m+\phi+(1-\alpha)/2]/\alpha}
  e^{-{M \omega r^{2}}/{2}}
  U\left(d,1+\frac{m+\phi+(1-\alpha)/2}{\alpha},M \omega r^{2}\right).
\label{eq:newsol}
\end{equation}
In order to guarantee that
$\chi(r)\in L^{2}(\mathbb{R},rdr)$, it is
advisable to study their behavior as $r\to 0$, which
implies analyzing the possible self-adjoint extensions.
Now, to construct the self-adjoint extensions, we must find the
deficiency subspaces,
\begin{equation}
[H_{0}+M^{2}\omega^{2}r^{2}]^{\dagger}\chi_{\pm}(r)=\pm i k_{0}^{2}\chi_{\pm}(r).
\label{eq:sefa}
\end{equation}
The solution to this equation is
\begin{equation}
\chi_{\pm}(r)=r^{[m+\phi+(1-\alpha)/2]/\alpha}e^{-{M \omega r^{2}}/{2}}
U\left(d_{\pm},1+\frac{m+\phi+(1-\alpha)/2}{\alpha},M \omega r^{2}\right),
\label{eq:phi}
\end{equation}
where
\begin{equation}
d_{\pm}=\frac{1}{2}\left(1+\frac{m+\phi+(1-\alpha)/2}{\alpha}\right)
  \mp \frac{i k_{0}}{2\omega}.
\end{equation}

Now considering the asymptotic behavior of $U(a,b,z)$ as $z\to 0$
\cite{Book.1972.Abramowitz}, let us find under which condition the
term,
\begin{equation}
\int |\chi_{\pm}(r)|^{2}rdr,  \label{eq:intc}
\end{equation}
has a finite contribution near the origin region.
Using Eq. \eqref{eq:phi} we found
\begin{equation}
  \lim_{r\to 0}|\chi_{\pm}(r)|^{2}r^{1+2[m+\phi+(1-\alpha)/2]/\alpha}
\to
  \left[\mathcal{A}_{1}r^{1+2[m+\phi+(1-\alpha)/2]/\alpha}+
    \mathcal{A}_{2}r^{1-2[m+\phi+(1-\alpha)/2]/\alpha}\right] ,
  \label{eq:lim}
\end{equation}
where $\mathcal{A}_{1}$and $\mathcal{A}_{2}$ are constants.
Studying Eq. \eqref{eq:lim}, we see that $\chi_{\pm}(r)$ is
square-integrable only for $[m+\phi+(1-\alpha)/2]/\alpha\in (-1,1)$.
In this case, since $N_{+}$ is expanded by $\chi_{+}(r)$ only, we have
that its dimension is $n_{+}=1$.
The same applies to $N_{-}$ and $\chi_{-}(r),$ resulting in $n_{-}=1$ .
Then, the Hilbert space, for both cases of
Eq. \eqref{eq:jrange}, contains vectors of the form
\begin{align}
  \chi_{\vartheta}(r)= {} &
  \chi_{m}(r) +c
  \left\{
    r^{|m+\phi+(1-\alpha)/2|/\alpha} e^{-{M \omega r^{2}}/{2}}
    U\left(d_{+},1+\frac{|m+\phi+(1-\alpha)/2|}{\alpha},
      M \omega r^{2}\right)\right. \nonumber \\
   & + \left. e^{i\vartheta}
    r^{|m+\phi+(1-\alpha)/2|/\alpha} e^{-{M \omega r^{2}}/{2}}
    U\left(d_{-},1+\frac{|m+\phi+(1-\alpha)/2|}{\alpha},
      M \omega r^{2}\right)
  \right\},
  \label{eq:solnew}
\end{align}
where $c$ is an arbitrary complex number,
$\chi_{m}(0)=\dot{\chi}_{m}(0)=0$ and
$\chi_{m}(r)\in L^{2}(\mathbb{R}^{+},rdr)$.
For a range of $\vartheta$, the behavior of the wave functions
\eqref{eq:solnew} was addressed in  \cite{JMP.1989.30.1053}.
The boundary condition at the origin will select the value of this
parameter.
The difference here is the presence of the harmonic term.
However, this harmonic term does not contribute to the BG logarithmic
derivative boundary condition (cf. Eq. \eqref{eq:logder}), since the
integration of the harmonic vanishes as $r_0\to 0^{+}$.
After this identification, proceeding in an analogous way we did in the
Section \ref{subsec:KS_method}, it is found that the
bound state energy is implicitly determined by the equation
\begin{align}
  \label{eq:energy_KS_HO}
  \frac
  {\Gamma\left({1/2+|m+\phi+(1-\alpha)/2|/2\alpha}-
      {E_{b}}/{2\omega}\right)}
  {\Gamma\left({1/2-|m+\phi+(1-\alpha)/2|/2\alpha}-
      {E_{b}}/{2\omega}\right)} = {} &
  -\frac{1}{(M \omega)^{|m+\phi+(1-\alpha)/2|/\alpha} r_{0}^{2|m+\phi+(1-\alpha)/2|/\alpha}}
  \left(
    \frac
    {\phi s+|m+\phi+(1-\alpha)/2|}
    {\phi s-|m+\phi+(1-\alpha)/2|}
  \right) \nonumber \\
  & \times
  \frac
  {\Gamma\left(1+|m+\phi+(1-\alpha)/2|/\alpha\right)}
  {\Gamma\left(1-|m+\phi+(1-\alpha)/2|/\alpha\right)}.
\end{align}
The above expression is too complicated to evaluate
the bound state energy, but its limiting features are
interesting.
If we take limit $r_{0}\to 0$ in this expression, the bound
state energy  are determined by the poles of the gamma
functions, i.e.,
\begin{align}
-1<\frac{m+\phi+(1-\alpha)/2}{\alpha}<0,&\qquad
E_{b}=\left(2n+1-\frac{|m+\phi+(1-\alpha)/2|}{\alpha}\right)\omega, \\
 0<\frac{m+\phi+(1-\alpha)/2}{\alpha}<1,&\qquad
E_{b}=\left(2n+1+\frac{|m+\phi+(1-\alpha)/2|}{\alpha}\right)\omega,
\end{align}
or
\begin{equation}
E_{b}=\left(2n+1\pm\frac{|m+\phi+(1-\alpha)/2|}{\alpha}\right)\omega,
\qquad n=0,1,2,\ldots.
\label{eq:landau+}
\end{equation}
The $+$ ($-$) sign refers to solutions which are regular
(irregular) at the origin.
This result coincide with the Eq. (1) of
Ref. \cite{PRL.1990.64.709}, for the special case of $\alpha=1$.
Another interesting case is that of vanishing HO potential.
This is achieved using the asymptotic behavior of the ratio
of gamma functions for $\omega\to 0$ \cite{Book.2010.NIST},
\begin{equation}
  \frac
  {\Gamma\left({1/2+|m+\phi+(1-\alpha)/2|/2\alpha}-
      {E}/{2 \omega}\right)}
  {\Gamma\left({1/2-|m+\phi+(1-\alpha)/2|/2\alpha}-
     {E}/{2 \omega}\right)}
  \sim
  \left(-\frac{E}{2 \omega}\right)^{|m+\phi+(1-\alpha)/2|/\alpha},
\label{eq:omegalimit}
\end{equation}
which holds for $E<0$ and this is the necessary condition
for the usual AB system have a bound state.
Using this limit
in the Eq. \eqref{eq:energy_KS_HO}, one finds
\begin{equation}
  E_{b}=-\frac{2}{Mr_{0}^{2}}
  \left[
    \left(
      \frac
      {\phi s+|m+\phi+(1-\alpha)/2|}
      {\phi s-|m+\phi+(1-\alpha)/2|}
    \right)
    \frac
    {\Gamma(1+|m+\phi+(1-\alpha)/2|/\alpha)}
    {\Gamma(1-|m+\phi+(1-\alpha)/2|/\alpha)}
  \right]^{\alpha/|m+\phi+(1-\alpha)/2|},
  \label{eq:energy_pure_AB}
\end{equation}
in agreement with the result obtained in
Eq. \eqref{eq:energy_KS}.
Thus, in the limit of vanishing harmonic oscillator, we recover
the usual AB problem in conical space, as it should be.

Now we have to remark that this result contains a subtlety that
must be interpreted as follows: the presence of the singularity
in the problem establishes the range 
$|m+\phi+(1-\alpha)/2|/\alpha<1$.
If we ignore the singularity and impose that the wave function
is regular at the origin ($\chi_{m}(r)\equiv\dot{\chi}_{m}(r)\equiv 0$),
we achieve the same spectrum of Eq. \eqref{eq:landau+}, but with
$[m+\phi+(1-\alpha)/2]/\alpha$ assuming any value
\cite{PLA.1994.195.90,JPA.2000.33.5513,EPL.1999.45.279}.
In this sense the self-adjoint extension prevents us from
obtaining a spectrum incompatible with the singular nature of
the Hamiltonian when we take into account the singular $\delta$
function \cite{PRA.2008.77.036101,PRD.1996.53.6829}.
We have to take into account that the true boundary condition is
that the wave function must be square-integrable through all
space, regardless it is irregular or regular at the origin
\cite{CMP.1991.139.103,PRD.1996.53.6829}.

\subsection{Determination of self-adjoint extension parameter}
\label{sec:dsaep}

In this section the self-adjoint extension parameter will be determined
in terms of the physics of the problem.
For our intent, it is more convenient to write the solution in
Eq. \eqref{eq:ku1} for  $r\neq 0$ solely in terms of the
confluent hypergeometric function $M(a,b,z)$, as
\begin{align}\label{eq:general_sol_2_HO}
  \chi_{m}(r)
  = {} & a_{m} (M\omega)^{{1/2+|m+\phi+(1-\alpha)/2|/2\alpha}}
  r^{|m+\phi+(1-\alpha)/2|/\alpha}
  e^{-{M \omega r^{2}}/{2}}
  M\left(d,1+\frac{|m+\phi+(1-\alpha)/2|}{\alpha},M\omega r^{2}\right)
  \nonumber \\
  & + b_{m}(M\omega)^{{1/2-|m+\phi+(1-\alpha)/2|/2\alpha}}
  r^{-|m+\phi+(1-\alpha)/2|/\alpha}
  e^{-{M \omega r^{2}}/{2}}
  M\left(d,1-\frac{|m+\phi+(1-\alpha)/2|}{\alpha},M\omega r^{2}\right),
\end{align}
where $a_{m}$, $b_{m}$ are the coefficients of the regular and
singular solutions, respectively.
By implementing Eq. \eqref{eq:general_sol_2_HO} into the
boundary condition \eqref{eq:bc}, we derive the following
relation between the coefficients $a_{m}$ and $b_{m}$:
\begin{equation}
  \lambda_{m} a_{m}(M \omega)^{|m+\phi+(1-\alpha)/2|/\alpha}=b_{m}
  \bigg(
  1-\frac{\lambda_{m}\;E}{4(1-|m+\phi+(1-\alpha)/2|/\alpha)}
  \lim_{r\rightarrow 0^{+}}r^{2-2|m+\phi+(1-\alpha)/2|/\alpha}
  \bigg),  \label{eq:coef_rel_1_HO}
\end{equation}
where $\lambda_{m}$ is the self-adjoint extension parameter for
the spin 1/2 AB problem plus a two-dimensional HO.
In the above equation, the coefficient of $B_{m}$ diverges as
$\lim_{r\rightarrow 0^{+}}r^{2-2|m+\phi+(1-\alpha)/2|/\alpha}$, if
$|m+\phi+(1-\alpha)/2|/\alpha>1$.
Thus, $b_{m}$ must be zero for $|m+\phi+(1-\alpha)/2|/\alpha>1$, and the
condition for the occurrence of a singular solution is
$|m+\phi+(1-\alpha)/2|/\alpha<1$.
So, the presence of an irregular solution stems from the fact
the operator is not self-adjoint for $|m+\phi+(1-\alpha)/2|/\alpha<1$,
recasting the condition of non-self-adjointness of the previews
sections.

Applying the normalizability condition in the
Eq. \eqref{eq:general_sol_2_HO}, yields the relation
\begin{equation}
  b_{m}=
  -\frac
  {\Gamma^{(+)}}
  {\Gamma^{(-)}}
  \frac
  {\Gamma\left({1/2+|m+\phi+(1-\alpha)/2|/2\alpha}-
      {E}/{2\omega}\right)}
  {\Gamma\left({1/2-|m+\phi+(1-\alpha)/2|/2\alpha}-
      {E}/{2\omega}\right)}
  a_{m}.
\label{eq:coef_rel_2_HO}
\end{equation}
From Eq. \eqref{eq:coef_rel_1_HO}, for
$|m+\phi+(1-\alpha)/2|/\alpha<1$ we have
$b_{m}=\lambda_{m}(M \omega)^{|m+\phi+(1-\alpha)/2|/\alpha}a_{m}$ and
by using Eq. \eqref{eq:coef_rel_2_HO}, the bound state energy is
implicitly determined by the equation
\begin{equation}
  \frac
  {\Gamma\left({1/2+|m+\phi+(1-\alpha)/2|/2\alpha}-
      {E_{b}}/{2\omega}\right)}
  {\Gamma\left({1/2-|m+\phi+(1-\alpha)/2|/2\alpha}-
      {E_{b}}/{2\omega}\right)}=
  -\frac{1}{\lambda_{m}(M \omega)^{|m+\phi+(1-\alpha)/2|/\alpha}}
  \frac
  {\Gamma(1+|m+\phi+(1-\alpha)/2|/\alpha)}
  {\Gamma(1-|m+\phi+(1-\alpha)/2|/\alpha)}
  \label{eq:energy_BG_HO}
\end{equation}
This results coincides with Eq. (53) of
Ref. \cite{JMP.1995.36.5453} for $\alpha=1$,
and using the result in Eq. \eqref{eq:omegalimit}, it is easy to see
that in the limit of vanishing oscillator potential, the spectrum of the
usual AB is recovered (cf. Eq. \eqref{eq:energy_BG}).
By comparing Eq. \eqref{eq:energy_BG_HO} with
Eq. \eqref{eq:energy_KS_HO}, we find
\begin{equation}
  \frac{1}{\lambda_{m}}=\frac{2}{r_{0}^{2|m+\phi+(1-\alpha)/2|/\alpha}}
  \left(
    \frac
    {\phi s+|m+\phi+(1-\alpha)/2|}
    {\phi s-|m+\phi+(1-\alpha)/2|}
  \right).
  \label{eq:alpha_j}
\end{equation}
Then, the relation between the self-adjoint extension parameter
and the physics of the problem for the usual AB has the same
mathematical structure as for the AB plus HO.
However, we must observe that the self-adjoint extension
parameter is negative for the usual AB, confirming the
restriction of negative values of the self-adjoint extension
parameter made in \cite{JMP.1995.36.5453}, in such way we have
an attractive $\delta$-function.
It is a necessary  condition to have a bound state in the usual
AB system.

\section{Conclusions}
\label{sec:conclusion}

We have presented a general regularization procedure to address
systems endowed with a singular Hamiltonian (due to localized fields
sources or quantum confinement).
Using  the KS approach, the bound states were determined in
terms of the physics of the problem, in a very consistent way
and without any arbitrary parameter.
In the sequel, we employed the BG approach.
By comparing the results of these approaches, we have determined
an expression for the self-adjoint extension parameter for the
bound state problem, which coincides with the one for scattering
problem.
We have thus obtained the S matrix in terms of the physics of
the problem as well.
In this point, we remark that the important results of
Refs. \cite{PRA.2002.66.032118,PRD.1994.50.7715,PRD.1989.40.1346}
are given in terms of an arbitrary self-adjoint extension
parameter.
In our work this parameter was determined in terms of the
physics of the problem.
The outcomes obtained by Park are a particular case of our
results for a fixed value of the self-adjoint extension
parameter.
To our knowledge, it was not known in the literature an
expression for the bound state energies for the AB with a
defined self-adjoint extension parameter.
In Ref. \cite{PRD.2012.85.041701} this expression was presented
by the first time, whose details are derived here.

To illustrate the applicability of our approach to other
physical systems, we deal with the spin-1/2 AB problem in conical space
plus a two dimensional HO.
Two cases were considered: (i) without and (ii) with the
inclusion of the $\delta$ function potential in the nonrelativistic
Hamiltonian.
Even though we have obtained an equivalent mathematical
expression for both cases, it has been shown that, in
(i) $[m+\phi+(1-\alpha)/2]/\alpha$ can assume any value while in
(ii) it is in the range $|m+\phi+(1-\alpha)/2|/\alpha| < 1$.
In the first case, it is reasonable to impose that the wave
function vanish at the origin.
However, this condition does not give a correct description of
the problem in the $r=0$ region.
Therefore, the energy spectrum obtained in the second case is
physically acceptable.
The presence of the singularity establishes that the effective
angular momentum must obey the condition
$|m+\phi+(1-\alpha)/2|/\alpha < 1$ and implies that irregular
solutions must be taken into account in this range.
The only situation in which we can neglect the $\delta$ function
potential is that one in which one looks only for topological phases.
A natural extension of this work is the inclusion of the Coulomb
potential, which naturally appears in two-dimensional systems, such as
graphene \cite{RMP.2009.81.109} and anyons systems
\cite{PRL.1982.49.957,IJMP.1989.B3.1001}.
This will be reported elsewhere.

\section*{Acknowledgments}

The authors would like to thank M. M. Ferreira Jr.,
E. R. Bezerra de Mello  and F. Moraes for the critical reading of the
manuscript and for helpful discussions.
F. M Andrade acknowledges research grants by Funda\c{c}\~{a}o
Arauc\'{a}ria project no. 205/2013,
E. O. Silva acknowledges research grants by
CNPq-(Universal) project no. 484959/2011-5 and
M. Pereira acknowledges research grants by Funda\c{c}\~{a}o
Arauc\'{a}ria.

\bibliographystyle{model1a-num-names}

\begin{thebibliography}{96}
\expandafter\ifx\csname natexlab\endcsname\relax\def\natexlab#1{#1}\fi
\providecommand{\url}[1]{\texttt{#1}}
\providecommand{\href}[2]{#2}
\providecommand{\path}[1]{#1}
\providecommand{\DOIprefix}{doi:}
\providecommand{\ArXivprefix}{arXiv:}
\providecommand{\URLprefix}{URL: }
\providecommand{\Pubmedprefix}{pmid:}
\providecommand{\doi}[1]{\href{http://dx.doi.org/#1}{\path{#1}}}
\providecommand{\Pubmed}[1]{\href{pmid:#1}{\path{#1}}}
\providecommand{\bibinfo}[2]{#2}
\ifx\xfnm\relax \def\xfnm[#1]{\unskip,\space#1}\fi
\bibitem[{Aharonov and Bohm(1959)}]{PR.1959.115.485}
\bibinfo{author}{Y.~Aharonov}, \bibinfo{author}{D.~Bohm},
  \bibinfo{journal}{Phys. Rev.} \bibinfo{volume}{115} (\bibinfo{year}{1959})
  \bibinfo{pages}{485}. \DOIprefix\doi{10.1103/PhysRev.115.485}.
\bibitem[{Ehrenberg and Siday(1949)}]{PPSB.1949.62.8}
\bibinfo{author}{W.~Ehrenberg}, \bibinfo{author}{R.~E. Siday},
  \bibinfo{journal}{Proc. Phys. Soc. B} \bibinfo{volume}{62}
  (\bibinfo{year}{1949}) \bibinfo{pages}{8--}.
  \DOIprefix\doi{10.1088/0370-1301/62/1/303}.
\bibitem[{Kronig and Penney(1931)}]{PRSLA.1931.130.499}
\bibinfo{author}{R.~L. Kronig}, \bibinfo{author}{W.~G. Penney},
  \bibinfo{journal}{Proc. R. Soc. Lond. A} \bibinfo{volume}{130}
  (\bibinfo{year}{1931}) \bibinfo{pages}{499}.
  \DOIprefix\doi{10.1098/RSPA.1931.0019}.
\bibitem[{Nouri-Zonoz and Parvizi(2013)}]{PRD.2013.88.023004}
\bibinfo{author}{M.~Nouri-Zonoz}, \bibinfo{author}{A.~Parvizi},
  \bibinfo{journal}{Phys. Rev. D} \bibinfo{volume}{88} (\bibinfo{year}{2013})
  \bibinfo{pages}{023004--}. \DOIprefix\doi{10.1103/PhysRevD.88.023004}.
\bibitem[{Vilenkin and Shellard(2000)}]{Book.2000.Vilenkin}
\bibinfo{author}{A.~Vilenkin}, \bibinfo{author}{E.~P.~S. Shellard},
  \bibinfo{title}{Cosmic Strings and Other Topological Defects},
  \bibinfo{publisher}{Cambridge University Pres}, \bibinfo{address}{Canbridge},
  \bibinfo{year}{2000}.
\bibitem[{Chu et~al.(2010)Chu, Mathur, and Vachaspati}]{PRD.2010.82.063515}
\bibinfo{author}{Y.-Z. Chu}, \bibinfo{author}{H.~Mathur},
  \bibinfo{author}{T.~Vachaspati}, \bibinfo{journal}{Phys. Rev. D}
  \bibinfo{volume}{82} (\bibinfo{year}{2010}) \bibinfo{pages}{063515--}.
  \DOIprefix\doi{10.1103/PhysRevD.82.063515}.
\bibitem[{de~Sousa~Gerbert(1989)}]{PRD.1989.40.1346}
\bibinfo{author}{P.~de~Sousa~Gerbert}, \bibinfo{journal}{Phys. Rev. D}
  \bibinfo{volume}{40} (\bibinfo{year}{1989}) \bibinfo{pages}{1346}.
  \DOIprefix\doi{10.1103/PhysRevD.40.1346}.
\bibitem[{Bezerra(2006)}]{BJP.2006.36.141}
\bibinfo{author}{V.~B. Bezerra}, \bibinfo{journal}{Braz. J. Phys.}
  \bibinfo{volume}{36} (\bibinfo{year}{2006}) \bibinfo{pages}{141 -- 156}.
  \DOIprefix\doi{10.1590/S0103-97332006000200006}.
\bibitem[{Hohensee et~al.(2012)Hohensee, Estey, Hamilton, Zeilinger, and
  M\"{u}ller}]{PRL.2012.108.230404}
\bibinfo{author}{M.~A. Hohensee}, \bibinfo{author}{B.~Estey},
  \bibinfo{author}{P.~Hamilton}, \bibinfo{author}{A.~Zeilinger},
  \bibinfo{author}{H.~M\"{u}ller}, \bibinfo{journal}{Phys. Rev. Lett.}
  \bibinfo{volume}{108} (\bibinfo{year}{2012}) \bibinfo{pages}{230404--}.
  \DOIprefix\doi{10.1103/PhysRevLett.108.230404}.
\bibitem[{Bezerra(1990)}]{AoP.1990.203.392}
\bibinfo{author}{V.~Bezerra}, \bibinfo{journal}{Ann. Phys. (NY)}
  \bibinfo{volume}{203} (\bibinfo{year}{1990}) \bibinfo{pages}{392 -- 409}.
  \DOIprefix\doi{10.1016/0003-4916(90)90175-N}.
\bibitem[{Bezerra(1997)}]{JMP.1997.38.2553}
\bibinfo{author}{V.~B. Bezerra}, \bibinfo{journal}{J. Math. Phys.}
  \bibinfo{volume}{38} (\bibinfo{year}{1997}) \bibinfo{pages}{2553--2564}.
  \DOIprefix\doi{10.1063/1.531995}.
\bibitem[{Furtado et~al.(2000)Furtado, B.~Bazerra, and
  Moraes}]{MPLA.2000.15.253}
\bibinfo{author}{C.~Furtado}, \bibinfo{author}{V.~B.~Bazerra},
  \bibinfo{author}{F.~Moraes}, \bibinfo{journal}{Mod. Phys. Lett. A}
  \bibinfo{volume}{15} (\bibinfo{year}{2000}) \bibinfo{pages}{253}.
  \DOIprefix\doi{10.1142/S0217732300000244}.
\bibitem[{de~A.~Marques and Bezerra(2004)}]{MPLA.2004.19.49}
\bibinfo{author}{G.~de~A.~Marques}, \bibinfo{author}{V.~B. Bezerra},
  \bibinfo{journal}{Mod. Phys. Lett. A} \bibinfo{volume}{19}
  (\bibinfo{year}{2004}) \bibinfo{pages}{49}.
  \DOIprefix\doi{10.1142/S021773230401237X}.
\bibitem[{Aliev and Gal'tsov(1989)}]{AoP.1989.193.142}
\bibinfo{author}{A.~Aliev}, \bibinfo{author}{D.~Gal'tsov},
  \bibinfo{journal}{Ann. Phys.} \bibinfo{volume}{193} (\bibinfo{year}{1989})
  \bibinfo{pages}{142--165}. \DOIprefix\doi{10.1016/0003-4916(89)90355-2}.
\bibitem[{Roy and Roychoudhury(1995)}]{PLB.1995.359.339}
\bibinfo{author}{P.~Roy}, \bibinfo{author}{R.~Roychoudhury},
  \bibinfo{journal}{Phys. Lett. B} \bibinfo{volume}{359} (\bibinfo{year}{1995})
  \bibinfo{pages}{339 -- 342}. \DOIprefix\doi{10.1016/0370-2693(95)01079-6}.
\bibitem[{Andrade and Silva(2013)}]{PLB.2013.719.467}
\bibinfo{author}{F.~M. Andrade}, \bibinfo{author}{E.~O. Silva},
  \bibinfo{journal}{Phys. Lett. B} \bibinfo{volume}{719} (\bibinfo{year}{2013})
  \bibinfo{pages}{467--471}. \DOIprefix\doi{10.1016/j.physletb.2013.01.062}.
\bibitem[{Hagen(1990)}]{PRL.1990.64.503}
\bibinfo{author}{C.~R. Hagen}, \bibinfo{journal}{Phys. Rev. Lett.}
  \bibinfo{volume}{64} (\bibinfo{year}{1990}) \bibinfo{pages}{503}.
  \DOIprefix\doi{10.1103/PhysRevLett.64.503}.
\bibitem[{Hagen(1993)}]{PRD.1993.48.5935}
\bibinfo{author}{C.~R. Hagen}, \bibinfo{journal}{Phys. Rev. D}
  \bibinfo{volume}{48} (\bibinfo{year}{1993}) \bibinfo{pages}{5935}.
  \DOIprefix\doi{10.1103/PhysRevD.48.5935}.
\bibitem[{Hagen and Park(1996)}]{AoP.1996.251.45}
\bibinfo{author}{C.~R. Hagen}, \bibinfo{author}{D.~K. Park},
  \bibinfo{journal}{Ann. Phys. (NY)} \bibinfo{volume}{251}
  (\bibinfo{year}{1996}) \bibinfo{pages}{45}.
  \DOIprefix\doi{10.1006/aphy.1996.0106}.
\bibitem[{Correa et~al.(2010)Correa, Falomir, Jakubsk\'{y}, and
  Plyushchay}]{AoP.2010.325.2653}
\bibinfo{author}{F.~Correa}, \bibinfo{author}{H.~Falomir},
  \bibinfo{author}{V.~Jakubsk\'{y}}, \bibinfo{author}{M.~S. Plyushchay},
  \bibinfo{journal}{Ann. Phys. (NY)} \bibinfo{volume}{325}
  (\bibinfo{year}{2010}) \bibinfo{pages}{2653--2667}.
  \DOIprefix\doi{10.1016/j.aop.2010.06.005}.
\bibitem[{Jakubsk\'{y} et~al.(2010)Jakubsk\'{y}, Nieto, and
  Plyushchay}]{PLB.2010.692.51}
\bibinfo{author}{V.~Jakubsk\'{y}}, \bibinfo{author}{L.-M. Nieto},
  \bibinfo{author}{M.~S. Plyushchay}, \bibinfo{journal}{Phys. Lett. B}
  \bibinfo{volume}{692} (\bibinfo{year}{2010}) \bibinfo{pages}{51--56}.
  \DOIprefix\doi{10.1016/j.physletb.2010.07.014}.
\bibitem[{Slobodeniuk et~al.(2011)Slobodeniuk, Sharapov, and
  Loktev}]{PRB.2011.84.125306}
\bibinfo{author}{A.~O. Slobodeniuk}, \bibinfo{author}{S.~G. Sharapov},
  \bibinfo{author}{V.~M. Loktev}, \bibinfo{journal}{Phys. Rev. B}
  \bibinfo{volume}{84} (\bibinfo{year}{2011}) \bibinfo{pages}{125306}.
  \DOIprefix\doi{10.1103/PhysRevB.84.125306}.
\bibitem[{Slobodeniuk et~al.(2010)Slobodeniuk, Sharapov, and
  Loktev}]{PRB.2010.82.075316}
\bibinfo{author}{A.~O. Slobodeniuk}, \bibinfo{author}{S.~G. Sharapov},
  \bibinfo{author}{V.~M. Loktev}, \bibinfo{journal}{Phys. Rev. B}
  \bibinfo{volume}{82} (\bibinfo{year}{2010}) \bibinfo{pages}{075316--}.
  \DOIprefix\doi{10.1103/PhysRevB.82.075316}.
\bibitem[{Belich et~al.(2011)Belich, Silva, Ferreira~Jr., and
  Orlando}]{PRD.2011.83.125025}
\bibinfo{author}{H.~Belich}, \bibinfo{author}{E.~O. Silva},
  \bibinfo{author}{M.~M. Ferreira~Jr.}, \bibinfo{author}{M.~T.~D. Orlando},
  \bibinfo{journal}{Phys. Rev. D} \bibinfo{volume}{83} (\bibinfo{year}{2011})
  \bibinfo{pages}{125025--}. \DOIprefix\doi{10.1103/PhysRevD.83.125025}.
\bibitem[{Hashimoto and Iizuka(2010)}]{PRD.2010.82.105023}
\bibinfo{author}{K.~Hashimoto}, \bibinfo{author}{N.~Iizuka},
  \bibinfo{journal}{Phys. Rev. D} \bibinfo{volume}{82} (\bibinfo{year}{2010})
  \bibinfo{pages}{105023--}. \DOIprefix\doi{10.1103/PhysRevD.82.105023}.
\bibitem[{Dolan et~al.(2011)Dolan, Oliveira, and Crispino}]{PLB.2011.701.485}
\bibinfo{author}{S.~R. Dolan}, \bibinfo{author}{E.~S. Oliveira},
  \bibinfo{author}{L.~C. Crispino}, \bibinfo{journal}{Physics Letters B}
  \bibinfo{volume}{701} (\bibinfo{year}{2011}) \bibinfo{pages}{485--489}.
  \DOIprefix\doi{http://dx.doi.org/10.1016/j.physletb.2011.06.013}.
\bibitem[{Dmitriev et~al.(2010)Dmitriev, Gornyi, Kachorovskii, and
  Polyakov}]{PRL.2010.105.036402}
\bibinfo{author}{A.~P. Dmitriev}, \bibinfo{author}{I.~V. Gornyi},
  \bibinfo{author}{V.~Y. Kachorovskii}, \bibinfo{author}{D.~G. Polyakov},
  \bibinfo{journal}{Phys. Rev. Lett.} \bibinfo{volume}{105}
  (\bibinfo{year}{2010}) \bibinfo{pages}{036402--}.
  \DOIprefix\doi{10.1103/PhysRevLett.105.036402}.
\bibitem[{Schelter et~al.(2012)Schelter, Trauzettel, and
  Recher}]{PRL.2012.108.106603}
\bibinfo{author}{J.~Schelter}, \bibinfo{author}{B.~Trauzettel},
  \bibinfo{author}{P.~Recher}, \bibinfo{journal}{Physical Review Letters}
  \bibinfo{volume}{108} (\bibinfo{year}{2012}) \bibinfo{pages}{106603--}.
  \DOIprefix\doi{10.1103/PhysRevLett.108.106603}.
\bibitem[{Tanaka and Cheon(2010)}]{PRA.2010.82.022104}
\bibinfo{author}{A.~Tanaka}, \bibinfo{author}{T.~Cheon},
  \bibinfo{journal}{Phys. Rev. A} \bibinfo{volume}{82} (\bibinfo{year}{2010})
  \bibinfo{pages}{022104--}. \DOIprefix\doi{10.1103/PhysRevA.82.022104}.
\bibitem[{Baltateanu(2011)}]{PLA.2011.375.2952}
\bibinfo{author}{D.~Baltateanu}, \bibinfo{journal}{Phys. Lett. A}
  \bibinfo{volume}{375} (\bibinfo{year}{2011}) \bibinfo{pages}{2952--2957}.
  \DOIprefix\doi{10.1016/j.physleta.2011.06.033}.
\bibitem[{Anacleto et~al.(2013)Anacleto, Brito, and
  Passos}]{PRD.2013.87.125015}
\bibinfo{author}{M.~A. Anacleto}, \bibinfo{author}{F.~A. Brito},
  \bibinfo{author}{E.~Passos}, \bibinfo{journal}{Phys. Rev. D}
  \bibinfo{volume}{87} (\bibinfo{year}{2013}) \bibinfo{pages}{125015--}.
  \DOIprefix\doi{10.1103/PhysRevD.87.125015}.
\bibitem[{Anacleto et~al.(2012)Anacleto, Brito, and
  Passos}]{PRD.2012.86.125015}
\bibinfo{author}{M.~A. Anacleto}, \bibinfo{author}{F.~A. Brito},
  \bibinfo{author}{E.~Passos}, \bibinfo{journal}{Phys. Rev. D}
  \bibinfo{volume}{86} (\bibinfo{year}{2012}) \bibinfo{pages}{125015}.
  \DOIprefix\doi{10.1103/PhysRevD.86.125015}.
\bibitem[{Das et~al.(2011)Das, Falomir, Nieto, Gamboa, and
  M\'endez}]{PRD.2011.84.045002}
\bibinfo{author}{A.~Das}, \bibinfo{author}{H.~Falomir},
  \bibinfo{author}{M.~Nieto}, \bibinfo{author}{J.~Gamboa},
  \bibinfo{author}{F.~M\'endez}, \bibinfo{journal}{Phys. Rev. D}
  \bibinfo{volume}{84} (\bibinfo{year}{2011}) \bibinfo{pages}{045002}.
  \DOIprefix\doi{10.1103/PhysRevD.84.045002}.
\bibitem[{Falomir et~al.(2002)Falomir, Gamboa, Loewe, M\'{e}ndez, and
  Rojas}]{PRD.2002.66.045018}
\bibinfo{author}{H.~Falomir}, \bibinfo{author}{J.~Gamboa},
  \bibinfo{author}{M.~Loewe}, \bibinfo{author}{F.~M\'{e}ndez},
  \bibinfo{author}{J.~C. Rojas}, \bibinfo{journal}{Phys. Rev. D}
  \bibinfo{volume}{66} (\bibinfo{year}{2002}) \bibinfo{pages}{045018--}.
  \DOIprefix\doi{10.1103/PhysRevD.66.045018}.
\bibitem[{Albeverio et~al.(2004)Albeverio, Gesztesy, Hoegh-Krohn, and
  Holden}]{Book.2004.Albeverio}
\bibinfo{author}{S.~Albeverio}, \bibinfo{author}{F.~Gesztesy},
  \bibinfo{author}{R.~Hoegh-Krohn}, \bibinfo{author}{H.~Holden},
  \bibinfo{title}{Solvable Models in Quantum Mechanics},
  \bibinfo{edition}{second} ed., \bibinfo{publisher}{AMS Chelsea Publishing},
  \bibinfo{address}{Providence, RI}, \bibinfo{year}{2004}.
\bibitem[{Reed and Simon(1975)}]{Book.1975.Reed.II}
\bibinfo{author}{M.~Reed}, \bibinfo{author}{B.~Simon}, \bibinfo{title}{Methods
  of Modern Mathematical Physics. II. Fourier Analysis, Self-Adjointness.},
  \bibinfo{publisher}{Academic Press}, \bibinfo{address}{New York - London},
  \bibinfo{year}{1975}.
\bibitem[{Andrade et~al.(2012)Andrade, Silva, and Pereira}]{PRD.2012.85.041701}
\bibinfo{author}{F.~M. Andrade}, \bibinfo{author}{E.~O. Silva},
  \bibinfo{author}{M.~Pereira}, \bibinfo{journal}{Phys. Rev. D}
  \bibinfo{volume}{85} (\bibinfo{year}{2012}) \bibinfo{pages}{041701(R)}.
  \DOIprefix\doi{10.1103/PhysRevD.85.041701}.
\bibitem[{Kay and Studer(1991)}]{CMP.1991.139.103}
\bibinfo{author}{B.~S. Kay}, \bibinfo{author}{U.~M. Studer},
  \bibinfo{journal}{Commun. Math. Phys.} \bibinfo{volume}{139}
  (\bibinfo{year}{1991}) \bibinfo{pages}{103}.
  \DOIprefix\doi{10.1007/BF02102731}.
\bibitem[{Bulla and Gesztesy(1985)}]{JMP.1985.26.2520}
\bibinfo{author}{W.~Bulla}, \bibinfo{author}{F.~Gesztesy}, \bibinfo{journal}{J.
  Math. Phys.} \bibinfo{volume}{26} (\bibinfo{year}{1985})
  \bibinfo{pages}{2520}. \DOIprefix\doi{10.1063/1.526768}.
\bibitem[{Filgueiras et~al.(2012)Filgueiras, Silva, and
  Andrade}]{JMP.2012.53.122106}
\bibinfo{author}{C.~Filgueiras}, \bibinfo{author}{E.~O. Silva},
  \bibinfo{author}{F.~M. Andrade}, \bibinfo{journal}{J. Math. Phys.}
  \bibinfo{volume}{53} (\bibinfo{year}{2012}) \bibinfo{pages}{122106}.
  \DOIprefix\doi{10.1063/1.4770048}. \href{http://arxiv.org/abs/1205.1155}{\tt
  arXiv:1205.1155}.
\bibitem[{Silva and Andrade(2013)}]{EPL.2013.101.51005}
\bibinfo{author}{E.~O. Silva}, \bibinfo{author}{F.~M. Andrade},
  \bibinfo{journal}{Europhys. Lett.} \bibinfo{volume}{101}
  (\bibinfo{year}{2013}) \bibinfo{pages}{51005--}.
  \DOIprefix\doi{10.1209/0295-5075/101/51005}.
\bibitem[{Silva et~al.(2013)Silva, Andrade, Filgueiras, and
  Belich}]{EPJC.2013.73.2402}
\bibinfo{author}{E.~O. Silva}, \bibinfo{author}{F.~M. Andrade},
  \bibinfo{author}{C.~Filgueiras}, \bibinfo{author}{H.~Belich},
  \bibinfo{journal}{Eur. Phys. J. C} \bibinfo{volume}{73}
  (\bibinfo{year}{2013}) \bibinfo{pages}{2402}.
  \DOIprefix\doi{10.1140/epjc/s10052-013-2402-1}.
\bibitem[{Andrade et~al.(2013)Andrade, Silva, Prud\^{e}ncio, and
  Filgueiras}]{JPG.2013.40.075007}
\bibinfo{author}{F.~M. Andrade}, \bibinfo{author}{E.~O. Silva},
  \bibinfo{author}{T.~Prud\^{e}ncio}, \bibinfo{author}{C.~Filgueiras},
  \bibinfo{journal}{J. Phys. G: Nucl. Part. Phys.} \bibinfo{volume}{40}
  (\bibinfo{year}{2013}) \bibinfo{pages}{075007--}.
  \DOIprefix\doi{10.1088/0954-3899/40/7/075007}.
\bibitem[{Filgueiras and Moraes(2007)}]{PLA.2007.361.13}
\bibinfo{author}{C.~Filgueiras}, \bibinfo{author}{F.~Moraes},
  \bibinfo{journal}{Phys. Lett. A} \bibinfo{volume}{361} (\bibinfo{year}{2007})
  \bibinfo{pages}{13}. \DOIprefix\doi{10.1016/j.physleta.2006.09.030}.
\bibitem[{Sokolov and Starobinski(1977)}]{SPD.1977.22.312}
\bibinfo{author}{D.~D. Sokolov}, \bibinfo{author}{A.~A. Starobinski},
  \bibinfo{journal}{Sov. Phys. Dokl.} \bibinfo{volume}{22}
  (\bibinfo{year}{1977}) \bibinfo{pages}{312}.
\bibitem[{Bezerra~de Mello(2004)}]{JHEP.2004.2004.16}
\bibinfo{author}{E.~R. Bezerra~de Mello}, \bibinfo{journal}{J. High Energy
  Phys.} \bibinfo{volume}{2004} (\bibinfo{year}{2004}) \bibinfo{pages}{016--}.
  \DOIprefix\doi{10.1088/1126-6708/2004/06/016}.
\bibitem[{Bakke et~al.(2008)Bakke, Nascimento, and
  Furtado}]{PRD.2008.78.064012}
\bibinfo{author}{K.~Bakke}, \bibinfo{author}{J.~R. Nascimento},
  \bibinfo{author}{C.~Furtado}, \bibinfo{journal}{Phys. Rev. D}
  \bibinfo{volume}{78} (\bibinfo{year}{2008}) \bibinfo{pages}{064012--}.
  \DOIprefix\doi{10.1103/PhysRevD.78.064012}.
\bibitem[{Bakke and Furtado(2010)}]{AdP.2010.522.447}
\bibinfo{author}{K.~Bakke}, \bibinfo{author}{C.~Furtado},
  \bibinfo{journal}{Ann. Phys. (Berlin)} \bibinfo{volume}{522}
  (\bibinfo{year}{2010}) \bibinfo{pages}{447--455}.
  \DOIprefix\doi{10.1002/andp.201000043}.
\bibitem[{de~Vega(1978)}]{PRD.1978.18.2932}
\bibinfo{author}{H.~J. de~Vega}, \bibinfo{journal}{Phys. Rev. D}
  \bibinfo{volume}{18} (\bibinfo{year}{1978}) \bibinfo{pages}{2932--2944}.
  \DOIprefix\doi{10.1103/PhysRevD.18.2932}.
\bibitem[{Brandenberger et~al.(1988)Brandenberger, Davis, and
  Matheson}]{NPB.1988.307.909}
\bibinfo{author}{R.~H. Brandenberger}, \bibinfo{author}{A.-C. Davis},
  \bibinfo{author}{A.~M. Matheson}, \bibinfo{journal}{Nucl. Phys. B}
  \bibinfo{volume}{307} (\bibinfo{year}{1988}) \bibinfo{pages}{909--923}.
  \DOIprefix\doi{10.1016/0550-3213(88)90112-5}.
\bibitem[{Alford and Wilczek(1989)}]{PRL.1989.62.1071}
\bibinfo{author}{M.~G. Alford}, \bibinfo{author}{F.~Wilczek},
  \bibinfo{journal}{Phys. Rev. Lett.} \bibinfo{volume}{62}
  (\bibinfo{year}{1989}) \bibinfo{pages}{1071--1074}.
  \DOIprefix\doi{10.1103/PhysRevLett.62.1071}.
\bibitem[{Kowalski et~al.(2002)Kowalski, Podlaski, and
  Rembieli\'{n}ski}]{PRA.2002.66.032118}
\bibinfo{author}{K.~Kowalski}, \bibinfo{author}{K.~Podlaski},
  \bibinfo{author}{J.~Rembieli\'{n}ski}, \bibinfo{journal}{Phys. Rev. A}
  \bibinfo{volume}{66} (\bibinfo{year}{2002}) \bibinfo{pages}{032118}.
  \DOIprefix\doi{10.1103/PhysRevA.66.032118}.
\bibitem[{de~Oliveira(2008)}]{Book.2008.Oliveira}
\bibinfo{author}{C.~R. de~Oliveira}, \bibinfo{title}{Intermediate Spectral
  Theory and Quantum Dynamics}, \bibinfo{publisher}{Birkh\"{a}user Basel},
  \bibinfo{address}{Switzerland}, \bibinfo{year}{2008}.
  \DOIprefix\doi{10.1007/978-3-7643-8795-2}.
\bibitem[{Akhiezer and Glazman(1993)}]{Book.1993.Akhiezer}
\bibinfo{author}{N.~I. Akhiezer}, \bibinfo{author}{I.~M. Glazman},
  \bibinfo{title}{Theory of Linear Operators in Hilbert Space},
  \bibinfo{publisher}{Dover}, \bibinfo{address}{New York},
  \bibinfo{year}{1993}.
\bibitem[{de~Sousa~Gerbert and Jackiw(1989)}]{CMP.1989.124.229}
\bibinfo{author}{P.~de~Sousa~Gerbert}, \bibinfo{author}{R.~Jackiw},
  \bibinfo{journal}{Commun. Math. Phys.} \bibinfo{volume}{124}
  (\bibinfo{year}{1989}) \bibinfo{pages}{229}.
  \DOIprefix\doi{10.1007/BF01219196}.
\bibitem[{Jackiw(1995)}]{Book.1995.Jackiw}
\bibinfo{author}{R.~Jackiw}, \bibinfo{title}{Diverse topics in theoretical and
  mathematical physics}, Advanced Series in Mathematical Physics,
  \bibinfo{publisher}{World Scientific}, \bibinfo{address}{Singapore},
  \bibinfo{year}{1995}.
\bibitem[{Voropaev et~al.(1991)Voropaev, Galtsov, and Spasov}]{PLB.1991.267.91}
\bibinfo{author}{S.~Voropaev}, \bibinfo{author}{D.~Galtsov},
  \bibinfo{author}{D.~Spasov}, \bibinfo{journal}{Phys. Lett. B}
  \bibinfo{volume}{267} (\bibinfo{year}{1991}) \bibinfo{pages}{91--94}.
  \DOIprefix\doi{10.1016/0370-2693(91)90529-Y}.
\bibitem[{Bordag and Voropaev(1994)}]{PLB.1994.333.238}
\bibinfo{author}{M.~Bordag}, \bibinfo{author}{S.~Voropaev},
  \bibinfo{journal}{Phys. Lett. B} \bibinfo{volume}{333} (\bibinfo{year}{1994})
  \bibinfo{pages}{238--244}. \DOIprefix\doi{10.1016/0370-2693(94)91037-5}.
\bibitem[{Bordag and Voropaev(1993)}]{JPA.1993.26.7637}
\bibinfo{author}{M.~Bordag}, \bibinfo{author}{S.~Voropaev},
  \bibinfo{journal}{J. Phys. A} \bibinfo{volume}{26} (\bibinfo{year}{1993})
  \bibinfo{pages}{7637--}. \DOIprefix\doi{10.1088/0305-4470/26/24/032}.
\bibitem[{Park and Oh(1994)}]{PRD.1994.50.7715}
\bibinfo{author}{D.~K. Park}, \bibinfo{author}{J.~G. Oh},
  \bibinfo{journal}{Phys. Rev. D} \bibinfo{volume}{50} (\bibinfo{year}{1994})
  \bibinfo{pages}{7715}. \DOIprefix\doi{10.1103/PhysRevD.50.7715}.
\bibitem[{Park(1995)}]{JMP.1995.36.5453}
\bibinfo{author}{D.~K. Park}, \bibinfo{journal}{J. Math. Phys.}
  \bibinfo{volume}{36} (\bibinfo{year}{1995}) \bibinfo{pages}{5453}.
  \DOIprefix\doi{10.1063/1.531271}.
\bibitem[{Filgueiras and Moraes(2008)}]{AoP.2008.323.3150}
\bibinfo{author}{C.~Filgueiras}, \bibinfo{author}{F.~Moraes},
  \bibinfo{journal}{Ann. Phys. (NY)} \bibinfo{volume}{323}
  (\bibinfo{year}{2008}) \bibinfo{pages}{3150--3157}.
  \DOIprefix\doi{10.1016/j.aop.2008.08.002}.
\bibitem[{Filgueiras et~al.(2010)Filgueiras, Silva, Oliveira, and
  Moraes}]{AoP.2010.325.2529}
\bibinfo{author}{C.~Filgueiras}, \bibinfo{author}{E.~O. Silva},
  \bibinfo{author}{W.~Oliveira}, \bibinfo{author}{F.~Moraes},
  \bibinfo{journal}{Ann. Phys. (NY)} \bibinfo{volume}{325}
  (\bibinfo{year}{2010}) \bibinfo{pages}{2529}.
  \DOIprefix\doi{10.1016/j.aop.2010.05.012}.
\bibitem[{Gesztesy et~al.(1987)Gesztesy, Albeverio, Hoegh-Krohn, and
  Holden}]{crll.1987.380.87}
\bibinfo{author}{F.~Gesztesy}, \bibinfo{author}{S.~Albeverio},
  \bibinfo{author}{R.~Hoegh-Krohn}, \bibinfo{author}{H.~Holden},
  \bibinfo{journal}{J. Reine Angew. Math.} \bibinfo{volume}{380}
  (\bibinfo{year}{1987}) \bibinfo{pages}{87}.
  \DOIprefix\doi{10.1515/crll.1987.380.87}.
\bibitem[{Dabrowski and Stovicek(1998)}]{JMP.1998.39.47}
\bibinfo{author}{L.~Dabrowski}, \bibinfo{author}{P.~Stovicek},
  \bibinfo{journal}{J. Math. Phys.} \bibinfo{volume}{39} (\bibinfo{year}{1998})
  \bibinfo{pages}{47--62}. \DOIprefix\doi{10.1063/1.532307}.
\bibitem[{Adami and Teta(1998)}]{LMP.1998.43.43}
\bibinfo{author}{R.~Adami}, \bibinfo{author}{A.~Teta}, \bibinfo{journal}{Lett.
  Math. Phys.} \bibinfo{volume}{43} (\bibinfo{year}{1998})
  \bibinfo{pages}{43--54}. \DOIprefix\doi{10.1023/A:1007330512611}.
\bibitem[{Hagen(1990)}]{PRL.1990.64.2347}
\bibinfo{author}{C.~R. Hagen}, \bibinfo{journal}{Phys. Rev. Lett.}
  \bibinfo{volume}{64} (\bibinfo{year}{1990}) \bibinfo{pages}{2347--2349}.
  \DOIprefix\doi{10.1103/PhysRevLett.64.2347}.
\bibitem[{Hagen(1991)}]{IJMPA.1991.6.3119}
\bibinfo{author}{C.~R. Hagen}, \bibinfo{journal}{Int. J. Mod. Phys. A}
  \bibinfo{volume}{6} (\bibinfo{year}{1991}) \bibinfo{pages}{3119}.
  \DOIprefix\doi{10.1142/S0217751X91001520}.
\bibitem[{Sakurai and Napolitano(2011)}]{Book.2011.Sakurai}
\bibinfo{author}{J.~J. Sakurai}, \bibinfo{author}{J.~Napolitano},
  \bibinfo{title}{Modern Quantum Mechanics}, \bibinfo{edition}{2nd ed.} ed.,
  \bibinfo{publisher}{Addison-Wesley}, \bibinfo{year}{2011}.
\bibitem[{Audretsch et~al.(1995)Audretsch, Jasper, and
  Skarzhinsky}]{JPA.1995.28.2359}
\bibinfo{author}{J.~Audretsch}, \bibinfo{author}{U.~Jasper},
  \bibinfo{author}{V.~D. Skarzhinsky}, \bibinfo{journal}{J. Phys. A}
  \bibinfo{volume}{28} (\bibinfo{year}{1995}) \bibinfo{pages}{2359}.
  \DOIprefix\doi{10.1088/0305-4470/28/8/026}.
\bibitem[{Coutinho et~al.(1992)Coutinho, Nogami, and
  Fernando~Perez}]{PRA.1992.46.6052}
\bibinfo{author}{F.~A.~B. Coutinho}, \bibinfo{author}{Y.~Nogami},
  \bibinfo{author}{J.~Fernando~Perez}, \bibinfo{journal}{Phys. Rev. A}
  \bibinfo{volume}{46} (\bibinfo{year}{1992}) \bibinfo{pages}{6052}.
  \DOIprefix\doi{10.1103/PhysRevA.46.6052}.
\bibitem[{Aharonov and Casher(1979)}]{PRA.1979.19.2461}
\bibinfo{author}{Y.~Aharonov}, \bibinfo{author}{A.~Casher},
  \bibinfo{journal}{Phys. Rev. A} \bibinfo{volume}{19} (\bibinfo{year}{1979})
  \bibinfo{pages}{2461--2462}. \DOIprefix\doi{10.1103/PhysRevA.19.2461}.
\bibitem[{de~Oliveira and Pereira(2010)}]{JPA.2010.43.354011}
\bibinfo{author}{C.~R. de~Oliveira}, \bibinfo{author}{M.~Pereira},
  \bibinfo{journal}{J. Phys. A} \bibinfo{volume}{43} (\bibinfo{year}{2010})
  \bibinfo{pages}{354011}. \DOIprefix\doi{10.1088/1751-8113/43/35/354011}.
\bibitem[{Abramowitz and Stegun(1972)}]{Book.1972.Abramowitz}
\bibinfo{editor}{M.~Abramowitz}, \bibinfo{editor}{I.~A. Stegun} (Eds.),
  \bibinfo{title}{Handbook of Mathematical Functions}, \bibinfo{publisher}{New
  York: Dover Publications}, \bibinfo{year}{1972}.
\bibitem[{Ruijsenaars(1983)}]{AoP.1983.146.1}
\bibinfo{author}{S.~N.~M. Ruijsenaars}, \bibinfo{journal}{Ann. Phys. (NY)}
  \bibinfo{volume}{146} (\bibinfo{year}{1983}) \bibinfo{pages}{1}.
  \DOIprefix\doi{10.1016/0003-4916(83)90051-9}.
\bibitem[{Bennaceur et~al.(1999)Bennaceur, Dobaczewski, and
  Ploszajczak}]{PRC.1999.60.34308}
\bibinfo{author}{K.~Bennaceur}, \bibinfo{author}{J.~Dobaczewski},
  \bibinfo{author}{M.~Ploszajczak}, \bibinfo{journal}{Phys. Rev. C}
  \bibinfo{volume}{60} (\bibinfo{year}{1999}) \bibinfo{pages}{034308--}.
  \DOIprefix\doi{10.1103/PhysRevC.60.034308}.
\bibitem[{Goldhaber(1977)}]{PRD.1977.16.1815}
\bibinfo{author}{A.~S. Goldhaber}, \bibinfo{journal}{Phys. Rev. D}
  \bibinfo{volume}{16} (\bibinfo{year}{1977}) \bibinfo{pages}{1815--1827}.
  \DOIprefix\doi{10.1103/PhysRevD.16.1815}.
\bibitem[{Kazama et~al.(1977)Kazama, Yang, and Goldhaber}]{PRD.1977.15.2287}
\bibinfo{author}{Y.~Kazama}, \bibinfo{author}{C.~N. Yang},
  \bibinfo{author}{A.~S. Goldhaber}, \bibinfo{journal}{Phys. Rev. D}
  \bibinfo{volume}{15} (\bibinfo{year}{1977}) \bibinfo{pages}{2287--2299}.
  \DOIprefix\doi{10.1103/PhysRevD.15.2287}.
\bibitem[{Grossman(1983)}]{PRL.1983.50.464}
\bibinfo{author}{B.~Grossman}, \bibinfo{journal}{Phys. Rev. Lett.}
  \bibinfo{volume}{50} (\bibinfo{year}{1983}) \bibinfo{pages}{464--467}.
  \DOIprefix\doi{10.1103/PhysRevLett.50.464}.
\bibitem[{Ganoulis(1993)}]{PLB.1993.298.63}
\bibinfo{author}{N.~Ganoulis}, \bibinfo{journal}{Phys. Lett. B}
  \bibinfo{volume}{298} (\bibinfo{year}{1993}) \bibinfo{pages}{63--68}.
  \DOIprefix\doi{10.1016/0370-2693(93)91708-U}.
\bibitem[{Davis et~al.(1994)Davis, Martin, and Ganoulis}]{NPB.1994.419.323}
\bibinfo{author}{A.~Davis}, \bibinfo{author}{A.~Martin},
  \bibinfo{author}{N.~Ganoulis}, \bibinfo{journal}{Nucl. Phys. B}
  \bibinfo{volume}{419} (\bibinfo{year}{1994}) \bibinfo{pages}{323--340}.
  \DOIprefix\doi{10.1016/0550-3213(94)90045-0}.
\bibitem[{Coutinho and Perez(1994)}]{PRD.1994.49.2092}
\bibinfo{author}{F.~A.~B. Coutinho}, \bibinfo{author}{J.~F. Perez},
  \bibinfo{journal}{Phys. Rev. D} \bibinfo{volume}{49} (\bibinfo{year}{1994})
  \bibinfo{pages}{2092--2097}. \DOIprefix\doi{10.1103/PhysRevD.49.2092}.
\bibitem[{Araujo et~al.(2001)Araujo, Coutinho, and Perez}]{JPA.2001.34.8859}
\bibinfo{author}{V.~S. Araujo}, \bibinfo{author}{F.~A.~B. Coutinho},
  \bibinfo{author}{J.~F. Perez}, \bibinfo{journal}{J. Phys. A}
  \bibinfo{volume}{34} (\bibinfo{year}{2001}) \bibinfo{pages}{8859--}.
  \DOIprefix\doi{10.1088/0305-4470/34/42/310}.
\bibitem[{Bonneau et~al.(2001)Bonneau, Faraut, and Valent}]{AJP.2001.69.322}
\bibinfo{author}{G.~Bonneau}, \bibinfo{author}{J.~Faraut},
  \bibinfo{author}{G.~Valent}, \bibinfo{journal}{Am. J. Phys.}
  \bibinfo{volume}{69} (\bibinfo{year}{2001}) \bibinfo{pages}{322}.
  \DOIprefix\doi{10.1119/1.1328351}.
\bibitem[{Sakurai(1967)}]{Book.1967.Sakurai}
\bibinfo{author}{J.~J. Sakurai}, \bibinfo{title}{Advanced Quantum Mechanics},
  \bibinfo{publisher}{Addison Wesley}, \bibinfo{year}{1967}.
\bibitem[{Doebner et~al.(1989)Doebner, Elmers, and
  Heidenreich}]{JMP.1989.30.1053}
\bibinfo{author}{H.~D. Doebner}, \bibinfo{author}{H.~J. Elmers},
  \bibinfo{author}{W.~F. Heidenreich}, \bibinfo{journal}{J. Math. Phys.}
  \bibinfo{volume}{30} (\bibinfo{year}{1989}) \bibinfo{pages}{1053--1059}.
  \DOIprefix\doi{10.1063/1.528375}.
\bibitem[{Blum et~al.(1990)Blum, Hagen, and Ramaswamy}]{PRL.1990.64.709}
\bibinfo{author}{T.~Blum}, \bibinfo{author}{C.~R. Hagen},
  \bibinfo{author}{S.~Ramaswamy}, \bibinfo{journal}{Phys. Rev. Lett.}
  \bibinfo{volume}{64} (\bibinfo{year}{1990}) \bibinfo{pages}{709--712}.
  \DOIprefix\doi{10.1103/PhysRevLett.64.709}.
\bibitem[{Olver et~al.(2010)Olver, Lozier, Boisvert, and
  Clark}]{Book.2010.NIST}
\bibinfo{editor}{F.~W.~J. Olver}, \bibinfo{editor}{D.~W. Lozier},
  \bibinfo{editor}{R.~F. Boisvert}, \bibinfo{editor}{C.~W. Clark} (Eds.),
  \bibinfo{title}{NIST Handbook of Mathematical Functions},
  \bibinfo{publisher}{Cambridge University Press}, \bibinfo{year}{2010}.
\bibitem[{Furtado et~al.(1994)Furtado, da~Cunha, Moraes, de~Mello, and
  Bezzerra}]{PLA.1994.195.90}
\bibinfo{author}{C.~Furtado}, \bibinfo{author}{B.~G. da~Cunha},
  \bibinfo{author}{F.~Moraes}, \bibinfo{author}{E.~de~Mello},
  \bibinfo{author}{V.~Bezzerra}, \bibinfo{journal}{Phys. Lett. A}
  \bibinfo{volume}{195} (\bibinfo{year}{1994}) \bibinfo{pages}{90--94}.
  \DOIprefix\doi{10.1016/0375-9601(94)90432-4}.
\bibitem[{Furtado and Moraes(2000)}]{JPA.2000.33.5513}
\bibinfo{author}{C.~Furtado}, \bibinfo{author}{F.~Moraes}, \bibinfo{journal}{J.
  Phys. A} \bibinfo{volume}{33} (\bibinfo{year}{2000}) \bibinfo{pages}{5513--}.
  \DOIprefix\doi{10.1088/0305-4470/33/31/306}.
\bibitem[{Furtado and Moraes(1999)}]{EPL.1999.45.279}
\bibinfo{author}{C.~Furtado}, \bibinfo{author}{F.~Moraes},
  \bibinfo{journal}{Europhys. Lett.} \bibinfo{volume}{45}
  (\bibinfo{year}{1999}) \bibinfo{pages}{279--}.
  \DOIprefix\doi{10.1209/epl/i1999-00159-8}.
\bibitem[{Hagen(2008)}]{PRA.2008.77.036101}
\bibinfo{author}{C.~R. Hagen}, \bibinfo{journal}{Phys. Rev. A}
  \bibinfo{volume}{77} (\bibinfo{year}{2008}) \bibinfo{pages}{036101--}.
  \DOIprefix\doi{10.1103/PhysRevA.77.036101}.
\bibitem[{Allen et~al.(1996)Allen, Kay, and Ottewill}]{PRD.1996.53.6829}
\bibinfo{author}{B.~Allen}, \bibinfo{author}{B.~S. Kay}, \bibinfo{author}{A.~C.
  Ottewill}, \bibinfo{journal}{Phys. Rev. D} \bibinfo{volume}{53}
  (\bibinfo{year}{1996}) \bibinfo{pages}{6829}.
  \DOIprefix\doi{10.1103/PhysRevD.53.6829}.
\bibitem[{Castro~Neto et~al.(2009)Castro~Neto, Guinea, Peres, Novoselov, and
  Geim}]{RMP.2009.81.109}
\bibinfo{author}{A.~H. Castro~Neto}, \bibinfo{author}{F.~Guinea},
  \bibinfo{author}{N.~M.~R. Peres}, \bibinfo{author}{K.~S. Novoselov},
  \bibinfo{author}{A.~K. Geim}, \bibinfo{journal}{Rev. Mod. Phys.}
  \bibinfo{volume}{81} (\bibinfo{year}{2009}) \bibinfo{pages}{109--162}.
  \DOIprefix\doi{10.1103/RevModPhys.81.109}.
\bibitem[{Wilczek(1982)}]{PRL.1982.49.957}
\bibinfo{author}{F.~Wilczek}, \bibinfo{journal}{Phys. Rev. Lett.}
  \bibinfo{volume}{49} (\bibinfo{year}{1982}) \bibinfo{pages}{957--959}.
  \DOIprefix\doi{10.1103/PhysRevLett.49.957}.
\bibitem[{Chen et~al.(1989)Chen, Wilczek, Witten, and
  Halperin}]{IJMP.1989.B3.1001}
\bibinfo{author}{Y.-H. Chen}, \bibinfo{author}{F.~Wilczek},
  \bibinfo{author}{E.~Witten}, \bibinfo{author}{B.~I. Halperin},
  \bibinfo{journal}{Int. J. Mod. Phys.} \bibinfo{volume}{B3}
  (\bibinfo{year}{1989}) \bibinfo{pages}{1001}.
  \DOIprefix\doi{10.1142/S0217979289000725}.

\end{thebibliography}

\end{document}